\documentclass[twocolumn,aps,prc,superscriptaddress,longbibliography,nofootinbib,showpacs,floatfix]{revtex4-1}

\usepackage{graphicx}	
\usepackage{xspace}     
\usepackage{color}
\usepackage{subfigure}
\newcommand{\pt}{\mbox{$p_T$}\xspace}
\newcommand{\mt}{\mbox{$m_T$}\xspace}

\newcommand{\Ncoll}{\mbox{$N_{\rm coll}$}\xspace}

\newcommand{\sqsn}{\mbox{$\sqrt{s_{_{NN}}}$}\xspace}
\newcommand{\ee}  {\mbox{$e^+e^-$}\xspace}
\newcommand{\mee}  {\mbox{$m_{e^+e^-}$}\xspace}
\newcommand{\dA}  {\mbox{$d$$+$Au}\xspace}
\newcommand{\pdA}  {\mbox{$p(d)$$+$A}\xspace}

\newcommand{\Aa}  {\mbox{$A$$+$$A$}\xspace}
\newcommand{\Au}  {\mbox{Au$+$Au}\xspace}
\newcommand{\pp}  {\mbox{$p$$+$$p$}\xspace}

\newcommand{\pythia}{{\sc pythia}\xspace}
\newcommand{\mcnlo}{{\sc mc@nlo}\xspace}
\newcommand{\powheg}{{\sc powheg}\xspace}
\newcommand{\herwig}{{\sc herwig}\xspace}

\newcommand{\cc}{\mbox{$c\bar{c}$}\xspace}
\newcommand{\bb}{\mbox{$b\bar{b}$}\xspace}
\newcommand{\qq}{\mbox{$q\bar{q}$}\xspace}
\newcommand{\Nunlike}{\mbox{$N_{+-}$}\xspace}
\newcommand{\Nlike}{\mbox{$N_{\pm\pm}$}\xspace}
\newcommand{\Blike}{\mbox{$B^{\rm comb}_{\pm\pm}$}\xspace}
\newcommand{\Bunlike}{\mbox{$B^{\rm comb}_{+-}$}\xspace}

\newcommand{\signal}{\mbox{$S_{+-}$}\xspace}
\newcommand{\dndy}{\mbox{$dN/dy$}\xspace}
\newcommand{\Rda}{\mbox{$R_{dAu}$}\xspace}

\begin{document}


\title{Measurements of $e^+e^-$ pairs from open heavy flavor in $p$+$p$ and  
$d$+$A$ collisions at $\sqrt{s_{_{NN}}}=200$ GeV}
\date{\today}

\newcommand{\abilene}{Abilene Christian University, Abilene, Texas 79699, USA}
\newcommand{\acadsin}{Institute of Physics, Academia Sinica, Taipei 11529, Taiwan}
\newcommand{\augie}{Department of Physics, Augustana University, Sioux Falls, South Dakota 57197, USA}
\newcommand{\banaras}{Department of Physics, Banaras Hindu University, Varanasi 221005, India}
\newcommand{\barc}{Bhabha Atomic Research Centre, Bombay 400 085, India}
\newcommand{\baruch}{Baruch College, City University of New York, New York, New York, 10010 USA}
\newcommand{\bnlcoll}{Collider-Accelerator Department, Brookhaven National Laboratory, Upton, New York 11973-5000, USA}
\newcommand{\bnlphys}{Physics Department, Brookhaven National Laboratory, Upton, New York 11973-5000, USA}
\newcommand{\caucr}{University of California-Riverside, Riverside, California 92521, USA}
\newcommand{\charlesczech}{Charles University, Ovocn\'{y} trh 5, Praha 1, 116 36, Prague, Czech Republic}
\newcommand{\chonbuk}{Chonbuk National University, Jeonju, 561-756, Korea}
\newcommand{\ciae}{Science and Technology on Nuclear Data Laboratory, China Institute of Atomic Energy, Beijing 102413, People's Republic of China}
\newcommand{\cns}{Center for Nuclear Study, Graduate School of Science, University of Tokyo, 7-3-1 Hongo, Bunkyo, Tokyo 113-0033, Japan}
\newcommand{\colorado}{University of Colorado, Boulder, Colorado 80309, USA}
\newcommand{\columbia}{Columbia University, New York, New York 10027 and Nevis Laboratories, Irvington, New York 10533, USA}
\newcommand{\czechtech}{Czech Technical University, Zikova 4, 166 36 Prague 6, Czech Republic}
\newcommand{\dapnia}{Dapnia, CEA Saclay, F-91191, Gif-sur-Yvette, France}
\newcommand{\debrecen}{Debrecen University, H-4010 Debrecen, Egyetem t{\'e}r 1, Hungary}
\newcommand{\elte}{ELTE, E{\"o}tv{\"o}s Lor{\'a}nd University, H-1117 Budapest, P{\'a}zm{\'a}ny P.~s.~1/A, Hungary}
\newcommand{\eszterhazy}{Eszterh\'azy K\'aroly University, K\'aroly R\'obert Campus, H-3200 Gy\"ngy\"os, M\'atrai \'ut 36, Hungary}
\newcommand{\ewha}{Ewha Womans University, Seoul 120-750, Korea}
\newcommand{\fit}{Florida Institute of Technology, Melbourne, Florida 32901, USA}
\newcommand{\fsu}{Florida State University, Tallahassee, Florida 32306, USA}
\newcommand{\gsu}{Georgia State University, Atlanta, Georgia 30303, USA}
\newcommand{\hiroshima}{Hiroshima University, Kagamiyama, Higashi-Hiroshima 739-8526, Japan}
\newcommand{\howard}{Department of Physics and Astronomy, Howard University, Washington, DC 20059, USA}
\newcommand{\ihepprot}{IHEP Protvino, State Research Center of Russian Federation, Institute for High Energy Physics, Protvino, 142281, Russia}
\newcommand{\illuiuc}{University of Illinois at Urbana-Champaign, Urbana, Illinois 61801, USA}
\newcommand{\inrras}{Institute for Nuclear Research of the Russian Academy of Sciences, prospekt 60-letiya Oktyabrya 7a, Moscow 117312, Russia}
\newcommand{\instpasczech}{Institute of Physics, Academy of Sciences of the Czech Republic, Na Slovance 2, 182 21 Prague 8, Czech Republic}
\newcommand{\isu}{Iowa State University, Ames, Iowa 50011, USA}
\newcommand{\jaea}{Advanced Science Research Center, Japan Atomic Energy Agency, 2-4 Shirakata Shirane, Tokai-mura, Naka-gun, Ibaraki-ken 319-1195, Japan}
\newcommand{\jinrdubna}{Joint Institute for Nuclear Research, 141980 Dubna, Moscow Region, Russia}
\newcommand{\jyvaskyla}{Helsinki Institute of Physics and University of Jyv{\"a}skyl{\"a}, P.O.Box 35, FI-40014 Jyv{\"a}skyl{\"a}, Finland}
\newcommand{\kek}{KEK, High Energy Accelerator Research Organization, Tsukuba, Ibaraki 305-0801, Japan}
\newcommand{\korea}{Korea University, Seoul, 136-701, Korea}
\newcommand{\kurchatov}{National Research Center ``Kurchatov Institute", Moscow, 123098 Russia}
\newcommand{\kyoto}{Kyoto University, Kyoto 606-8502, Japan}
\newcommand{\labllr}{Laboratoire Leprince-Ringuet, Ecole Polytechnique, CNRS-IN2P3, Route de Saclay, F-91128, Palaiseau, France}
\newcommand{\lahorelums}{Physics Department, Lahore University of Management Sciences, Lahore 54792, Pakistan}
\newcommand{\lawllnl}{Lawrence Livermore National Laboratory, Livermore, California 94550, USA}
\newcommand{\losalamos}{Los Alamos National Laboratory, Los Alamos, New Mexico 87545, USA}
\newcommand{\lpc}{LPC, Universit{\'e} Blaise Pascal, CNRS-IN2P3, Clermont-Fd, 63177 Aubiere Cedex, France}
\newcommand{\lund}{Department of Physics, Lund University, Box 118, SE-221 00 Lund, Sweden}
\newcommand{\maryland}{University of Maryland, College Park, Maryland 20742, USA}
\newcommand{\mass}{Department of Physics, University of Massachusetts, Amherst, Massachusetts 01003-9337, USA}
\newcommand{\michigan}{Department of Physics, University of Michigan, Ann Arbor, Michigan 48109-1040, USA}
\newcommand{\muenster}{Institut f\"ur Kernphysik, University of Muenster, D-48149 Muenster, Germany}
\newcommand{\muhlenberg}{Muhlenberg College, Allentown, Pennsylvania 18104-5586, USA}
\newcommand{\myongji}{Myongji University, Yongin, Kyonggido 449-728, Korea}
\newcommand{\nagasaki}{Nagasaki Institute of Applied Science, Nagasaki-shi, Nagasaki 851-0193, Japan}
\newcommand{\nara}{Nara Women's University, Kita-uoya Nishi-machi Nara 630-8506, Japan}
\newcommand{\natmephi}{National Research Nuclear University, MEPhI, Moscow Engineering Physics Institute, Moscow, 115409, Russia}
\newcommand{\newmex}{University of New Mexico, Albuquerque, New Mexico 87131, USA}
\newcommand{\nmsu}{New Mexico State University, Las Cruces, New Mexico 88003, USA}
\newcommand{\ohio}{Department of Physics and Astronomy, Ohio University, Athens, Ohio 45701, USA}
\newcommand{\ornl}{Oak Ridge National Laboratory, Oak Ridge, Tennessee 37831, USA}
\newcommand{\orsay}{IPN-Orsay, Univ.~Paris-Sud, CNRS/IN2P3, Universit\'e Paris-Saclay, BP1, F-91406, Orsay, France}
\newcommand{\peking}{Peking University, Beijing 100871, People's Republic of China}
\newcommand{\pnpi}{PNPI, Petersburg Nuclear Physics Institute, Gatchina, Leningrad region, 188300, Russia}
\newcommand{\riken}{RIKEN Nishina Center for Accelerator-Based Science, Wako, Saitama 351-0198, Japan}
\newcommand{\rikjrbrc}{RIKEN BNL Research Center, Brookhaven National Laboratory, Upton, New York 11973-5000, USA}
\newcommand{\rikkyo}{Physics Department, Rikkyo University, 3-34-1 Nishi-Ikebukuro, Toshima, Tokyo 171-8501, Japan}
\newcommand{\saispbstu}{Saint Petersburg State Polytechnic University, St.~Petersburg, 195251 Russia}
\newcommand{\saopaulo}{Universidade de S{\~a}o Paulo, Instituto de F\'{\i}sica, Caixa Postal 66318, S{\~a}o Paulo CEP05315-970, Brazil}
\newcommand{\seoulnat}{Department of Physics and Astronomy, Seoul National University, Seoul 151-742, Korea}
\newcommand{\stonybrkc}{Chemistry Department, Stony Brook University, SUNY, Stony Brook, New York 11794-3400, USA}
\newcommand{\stonycrkp}{Department of Physics and Astronomy, Stony Brook University, SUNY, Stony Brook, New York 11794-3800, USA}
\newcommand{\subatech}{SUBATECH (Ecole des Mines de Nantes, CNRS-IN2P3, Universit{\'e} de Nantes) BP 20722-44307, Nantes, France}
\newcommand{\tenn}{University of Tennessee, Knoxville, Tennessee 37996, USA}
\newcommand{\titech}{Department of Physics, Tokyo Institute of Technology, Oh-okayama, Meguro, Tokyo 152-8551, Japan}
\newcommand{\tsukuba}{Center for Integrated Research in Fundamental Science and Engineering, University of Tsukuba, Tsukuba, Ibaraki 305, Japan}
\newcommand{\vandy}{Vanderbilt University, Nashville, Tennessee 37235, USA}
\newcommand{\waseda}{Waseda University, Advanced Research Institute for Science and Engineering, 17  Kikui-cho, Shinjuku-ku, Tokyo 162-0044, Japan}
\newcommand{\weizmann}{Weizmann Institute, Rehovot 76100, Israel}
\newcommand{\wigner}{Institute for Particle and Nuclear Physics, Wigner Research Centre for Physics, Hungarian Academy of Sciences (Wigner RCP, RMKI) H-1525 Budapest 114, POBox 49, Budapest, Hungary}
\newcommand{\yonsei}{Yonsei University, IPAP, Seoul 120-749, Korea}
\newcommand{\zagreb}{Department of Physics, Faculty of Science, University of Zagreb, Bijeni\v{c}ka 32, HR-10002 Zagreb, Croatia}
\affiliation{\abilene}
\affiliation{\acadsin}
\affiliation{\augie}
\affiliation{\banaras}
\affiliation{\barc}
\affiliation{\baruch}
\affiliation{\bnlcoll}
\affiliation{\bnlphys}
\affiliation{\caucr}
\affiliation{\charlesczech}
\affiliation{\chonbuk}
\affiliation{\ciae}
\affiliation{\cns}
\affiliation{\colorado}
\affiliation{\columbia}
\affiliation{\czechtech}
\affiliation{\dapnia}
\affiliation{\debrecen}
\affiliation{\elte}
\affiliation{\eszterhazy}
\affiliation{\ewha}
\affiliation{\fit}
\affiliation{\fsu}
\affiliation{\gsu}
\affiliation{\hiroshima}
\affiliation{\howard}
\affiliation{\ihepprot}
\affiliation{\illuiuc}
\affiliation{\inrras}
\affiliation{\instpasczech}
\affiliation{\isu}
\affiliation{\jaea}
\affiliation{\jinrdubna}
\affiliation{\jyvaskyla}
\affiliation{\kek}
\affiliation{\korea}
\affiliation{\kurchatov}
\affiliation{\kyoto}
\affiliation{\labllr}
\affiliation{\lahorelums}
\affiliation{\lawllnl}
\affiliation{\losalamos}
\affiliation{\lpc}
\affiliation{\lund}
\affiliation{\maryland}
\affiliation{\mass}
\affiliation{\michigan}
\affiliation{\muenster}
\affiliation{\muhlenberg}
\affiliation{\myongji}
\affiliation{\nagasaki}
\affiliation{\nara}
\affiliation{\natmephi}
\affiliation{\newmex}
\affiliation{\nmsu}
\affiliation{\ohio}
\affiliation{\ornl}
\affiliation{\orsay}
\affiliation{\peking}
\affiliation{\pnpi}
\affiliation{\riken}
\affiliation{\rikjrbrc}
\affiliation{\rikkyo}
\affiliation{\saispbstu}
\affiliation{\saopaulo}
\affiliation{\seoulnat}
\affiliation{\stonybrkc}
\affiliation{\stonycrkp}
\affiliation{\subatech}
\affiliation{\tenn}
\affiliation{\titech}
\affiliation{\tsukuba}
\affiliation{\vandy}
\affiliation{\waseda}
\affiliation{\weizmann}
\affiliation{\wigner}
\affiliation{\yonsei}
\affiliation{\zagreb}
\author{A.~Adare} \affiliation{\colorado} 
\author{S.~Afanasiev} \affiliation{\jinrdubna} 
\author{C.~Aidala} \affiliation{\mass} \affiliation{\michigan} 
\author{N.N.~Ajitanand} \affiliation{\stonybrkc} 
\author{Y.~Akiba} \email[PHENIX Spokesperson: ]{akiba@rcf.rhic.bnl.gov} \affiliation{\riken} \affiliation{\rikjrbrc} 
\author{H.~Al-Bataineh} \affiliation{\nmsu} 
\author{J.~Alexander} \affiliation{\stonybrkc} 
\author{M.~Alfred} \affiliation{\howard} 
\author{K.~Aoki} \affiliation{\kek} \affiliation{\kyoto} \affiliation{\riken} 
\author{N.~Apadula} \affiliation{\isu} \affiliation{\stonycrkp} 
% ...
\collaboration{PHENIX Collaboration} \noaffiliation


\begin{abstract}

We report a measurement of $e^+e^-$ pairs from semileptonic heavy-flavor 
decays in $p$$+$$p$ collisions at $\sqrt{s_{_{NN}}}=200$~GeV. The $e^+e^-$ 
pair yield from $b\bar{b}$ and $c\bar{c}$ is separated by exploiting a 
double differential fit done simultaneously in dielectron invariant mass 
and $p_T$. We used three different event generators, {\sc pythia}, 
{\sc mc@nlo}, and {\sc powheg}, to simulate the $e^+e^-$ spectra from 
$c\bar{c}$ and $b\bar{b}$ production. The data can be well described by 
all three generators within the detector acceptance. However, when 
using the generators to extrapolate to $4\pi$, significant differences 
are observed for the total cross section. These difference are less 
pronounced for $b\bar{b}$ than for $c\bar{c}$. The same model 
dependence was observed in already published $d$$+$$A$ data. The 
$p$$+$$p$ data are also directly compared with $d$$+$$A$ data in mass 
and $p_T$, and within the statistical accuracy no nuclear modification 
is seen.

\end{abstract}

	
\maketitle

\section{Introduction}

Heavy quarks such as charm and bottom are excellent probes to 
understand the properties of the quark gluon plasma (QGP) created in 
high energy heavy-ion collisions. Both charm and bottom quarks have 
masses significantly larger than the quantum chromodynamics (QCD) scale 
parameter $\Lambda_{QCD}\approx$ 0.2 GeV, and as such, their production 
is limited to the primordial nucleon-nucleon collisions. Heavy flavor 
production in the subsequent early, hot stages of heavy-ion collisions 
is not significant and thus any modification of the primordial heavy 
flavor phase space distributions in heavy ion collisions will be the 
result of the quarks traversing the QGP and later phases in the space 
time evolution.

Prior to the studies of heavy flavor production done at the Relativistic 
Heavy Ion Collider (RHIC), the high 
\pt suppression \cite{Adcox:2001jp,Adler:2003qi,Adams:2003kg} of light 
flavor hadrons was primarily associated to radiative energy loss via 
medium-induced gluon radiation. This predicted a distinctive mass 
hierarchy of high \pt suppression as measured via the nuclear 
modification factor $R_{AA}$, implying that hadrons with heavy flavor 
will have a smaller suppression: 
$R_{AA}^{\pi^0}<R_{AA}^{c}<R_{AA}^{b}$. $R_{AA}^{\pi^0}$ denotes the 
nuclear modification of $\pi^0$, defined as the ratio of yield measured 
in $AA$ collisions to the yield measured in \pp collisions scaled by 
the number of binary collisions for $AA$ system, and $R_{AA}^{c}$(or 
$R_{AA}^{b}$) denotes the same for charm (or bottom) quarks. However, 
the measurements showed similar suppression for light and heavy flavor 
hadrons. Including collisional energy loss via elastic scattering, 
which is more important for heavy flavor than for the light quarks, 
leads to a qualitative explanation of the large energy loss for heavy 
flavor ~\cite{He:2011qa,Sharma:2012dy}. But other approaches are 
similarly successful, including Langevin-based transport 
models~\cite{Moore:2004tg, vanHees:2005wb} and AdS/CFT (anti 
de-Sitter-space/conformal field theory) string drag energy loss 
models~\cite{Horowitz:2008ig}. Despite significant effort, a full 
quantitative understanding of the energy loss has not been achieved 
yet.

To test different theoretical approaches, it is crucial to understand 
primordial heavy flavor production, and any modifications there in the 
presence of nuclei. Primordial heavy flavor production can be studied 
in \pp collisions. When nuclei are involved in a collision, one might 
expect modifications to the initial state, which can be described as 
shadowing or anti shadowing of the parton distribution functions. Also 
modifications in the final state that can be expressed as changes of 
the fragmentation process are possible, for example, via energy loss or 
re-scattering in cold nuclear matter. It is commonly accepted that 
these effects are observable in \pdA collisions, where QGP formation is 
not expected. Differences between the single electron spectra from 
heavy flavor decays from \dA data and \pp data have been interpreted as 
cold nuclear matter effects~\cite{Adare:2012yxa}.

Recently hints of collectivity have been found in high multiplicity 
events from collisions of small nucleus with a large nuclei, which 
suggests that hot matter might even be formed in small systems. 
However, one would not expect sizable collective effects on the heavy 
flavor phase space distributions even if hot matter is created due to 
the small reaction volume in these collisions.

The primordial heavy flavor production can be calculated in the 
framework of perturbative QCD (pQCD).  Therefore, measurement of heavy 
flavor in \pp serves as a test for these calculations and can be used 
to improve Monte-Carlo (MC) generators. Results from MC generators can 
be scaled to \Aa or \pdA collision systems with the number of binary 
collisions and serve as a reference for observables in the absence of 
\pp data.

At RHIC, open heavy flavor production has been measured by both the 
PHENIX and STAR experiments in different collision systems, spanning 
\pp, \dA, Cu$+$Cu and \Au systems, and by exploiting various techniques 
such as single electrons/muons via semi-leptonic 
decays~\cite{Adare:2010de,Adare:2012yxa}, electron-hadron 
correlations~\cite{Aggarwal:2010xp}, $e-\mu$~\cite{Adare:2013xlp}, 
\ee~\cite{Adare:2014iwg} and also via reconstruction of 
$D$-mesons~\cite{Adamczyk:2014uip}. This paper reports the measurement 
of heavy flavor production via dielectrons in \pp collisions at 
midrapidity. The \ee pairs coming from the semi-leptonic decays of 
charm and bottom dominate different regions in mass and \pt allowing to 
disentangle the two contributions. Studying the \ee pairs from heavy 
flavor may also provide sensitivity to the heavy quark correlations 
which is important to constrain the MC models. The results from the \pp 
data from this paper can be directly compared to the previously 
published \dA data~\cite{Adare:2014iwg} that exploited the same 
technique.

The paper is organized as follows: Section II describes the 
experimental apparatus and trigger. Section III details the data 
analysis including electron identification, background subtraction, and 
efficiency corrections. A description of the hadronic cocktail and 
heavy flavor generators is outlined in Section IV, followed by studies 
of systematic uncertainties in Section V. The data are presented as 
double differential spectra in mass and $p_T$ in Section VI. The final 
results and the comparison of \pp and \dA, as well as the comparison to 
several models of charm and bottom production are discussed in Section 
VII. Section VIII gives our summary and conclusions.

\section{Experiment}

A detailed description of the PHENIX detector is available 
in~\cite{Adcox:2003zm}. We focus here on the components of the two 
central arm spectrometers and the beam-beam counters (BBCs) that are 
critical for the analysis of \ee pairs. Each of the two central arms 
cover a pseudorapidity range of $|\eta|<$0.35 
(70$^\circ<\theta<$110$^\circ$) and 90$^\circ$ in azimuthal angle 
$\phi$. They are located almost back-to-back, with an angular gap of 
67.5$^\circ$ between them at the top. They span a range from about 220 
cm to 500 cm radially from the beam axis. The location of collision 
vertex in the beam direction, the collision time, and the minimum bias 
(MB) trigger are provided by a system of two beam-beam counters (BBC) 
that are located at a distance of 144 cm from the nominal interaction 
point on either side. Each BBC covers the full azimuth and a rapidity 
range of 3.1 $< |\eta| <$ 3.9. The collision vertex resolution in the 
beam direction is approximately 2 cm for \pp collisions. The MB trigger 
requires a coincidence between both sides with at least one hit on each 
side, and accepts the events if the BBC vertex is within 38 cm of the 
nominal interaction point. The BBC cross section in \pp collisions was 
determined via the van der Meer scan technique~\cite{Drees:2001kn} and 
was found to be $\sigma_{\rm BBC}^{\pp}~=~23.0~\pm~2.2$ mb or 0.545 
$\pm$ 0.06 of the total inelastic \pp cross section of 
$\sigma_{inel}^{\pp}~=~42~\pm~3$ mb.

There are two primary charged particle tracking subsystems in PHENIX: 
drift chambers (DC) and pad chambers (PC)~\cite{Adcox:2003zp}. The DC 
along with first layer of PC (PC1) form the inner tracking 
system used here. The DC measures the trajectories of charged particles 
in the plane perpendicular to the beams and allows one to determine 
their charge and transverse momentum \pt. The PC1 provides a space 
point along the trajectory of charged particles, which is used to 
determine the polar angle $\theta$ and $z$-coordinate of the track. The 
momentum resolution for this data set is $\delta p/p$ = 0.011 $\oplus$ 
0.0116$\,p$[GeV/$c$].

Each central arm is equipped with a ring imaging \v{C}erenkov (RICH) 
detector that serves as the primary device for electron identification. 
With $CO_2$ as a radiator gas, an $e/\pi$ rejection of better than one 
part in $\sim$10$^{3}$ is achieved, for the tracks with momenta below 
the pion \v{C}erenkov threshold of $\sim$ 4.87 GeV/$c$. For each 
electron on average 10 \v{C}erenkov photons are reconstructed on a ring 
of 11.8 cm diameter with an array of photo multiplier tubes. Further 
electron identification is provided by the electromagnetic calorimeters 
(EMCal) that measure the spatial position and energy of the electrons. 
This is achieved by placing a cut on the ratio of the energy measured 
by EMCal and momentum given by the DC~\cite{Adare:2010de}.

To select potentially interesting events containing electrons, PHENIX 
uses a hardware trigger known as ERT (EMCal-RICH) trigger. The trigger 
is based on the online sum of the energy signals in a tile of 
$2\times2$ EMCal towers \cite{Aphecetche:2003zr}. For all EMCal trigger 
tiles above a predetermined threshold value, the location of the EMCal 
tile is matched with hits in the corresponding RICH tile ($4\times5$ 
PMTs). The location of the RICH tile depends on the energy of the 
trigger particle and is determined from a look-up table, assuming that 
the trigger particle is an electron. If a spatial match is found, an 
ERT trigger is issued.

\section{Data Analysis}

The data reported in this paper were collected during the 2006 RHIC \pp 
run. The data were recorded with the PHENIX detector using a MB 
trigger and the ERT trigger. The ERT energy threshold $E_{th}$ was 
set to 400 MeV for majority of the run, but was raised to 600 MeV 
towards the end of the run. A total of 855 million ERT triggered events 
corresponding to 143 billion sampled MB events were analyzed.  The 
corresponding integrated luminosity is $6.6~p$b$^{-1}$.

\subsection{Event selection and electron identification cuts}

The \pp analysis described here is very similar to the analysis of \ee 
pairs from \dA collisions published in~\cite{Adare:2014iwg}. A detailed 
description of electron identification as well as pair cuts can be 
found in~\cite{Adare:2014iwg,phd_deepali}. Events selected were 
required to have a reconstructed $z$-vertex within 30 cm of the nominal 
interaction point. Charged tracks reconstructed using the DC and PC1 
must pass stringent quality cuts and an explicit cut of \pt $>$ 0.2 
GeV/$c$. The track is then selected as an electron if at least two 
photomultiplier tubes registered \v{C}erenkov photons on the expected 
ring. Additionally, electron tracks are required to have a good match 
to a cluster in EMCal, and the energy of the cluster must satisfy the 
requirement $E/p~>$ 0.5, where $p$ is the momentum measured by the DCs.

\subsection{Combining tracks to electron pairs}

All electron tracks in a given event are combined to form pairs. We 
apply a minimum cut on the transverse mass of the pair, 
$m_T = \sqrt{(mc)^2+p_T^2} >$ 650 MeV/$c$. For the data taken using an 
ERT trigger, we require that one of the tracks of the pair has a \pt of 
at least 500 (700) MeV/c exceeding the nominal energy threshold 400 (or 
600) MeV of the trigger.

These pairs can be subdivided into three groups: (i) Signal pairs that 
we want to extract. In \pp collisions these are mostly from the decays 
of pseudoscalar mesons, vector mesons, \and heavy flavor mesons. (ii) 
Combinatorial pairs, which are an undesired background. These result 
from the combinations of unrelated tracks in any given event, such as 
combining tracks that originate from two different decays. (iii) 
Correlated background pairs, which are also undesired, but these pairs 
do not result from random combinations of tracks. The combinatorial and 
correlated background pairs should be removed to extract the signal 
pairs. Most of this is done via a statistical subtraction discussed in 
detail in Section~\ref{sec:ee_spec}. However, some of the correlated 
background can be removed through cuts on the pairs referred to as pair 
cuts.

There are several sources of correlated pairs which are treated 
separately. One type of correlated pairs result from detector problems 
or ambiguities in the pattern recognition. The most important 
contributor are hadron tracks that are parallel to electron tracks in 
the RICH. Both tracks share the same ring and are identified as 
electrons. These pairs can be removed by placing a cut on the distance 
between the projections of both pairs to the RICH focal plane. Similar 
cuts to avoid detector overlaps are placed on all detector systems.

Another type of correlated pairs are the ones that originate from the 
photons that convert to \ee pairs in the detector material in front of 
the tracking detectors, e.g. in the beryllium beam pipe (0.3\% of a 
radiation length ($X_0$) for the year 2006). The tracks from these 
pairs get reconstructed with an incorrect momentum, because the tracking 
algorithm assumes that all tracks originate from the vertex and hence 
traverse the full magnetic field. This leads to an artificial opening 
angle of the pairs that is always oriented perpendicular to the axial 
magnetic field. A cut on the orientation of the opening angle removes 
these pairs from the sample.  See~\cite{Adare:2014iwg,Adare:2009qk} for 
a full description of the pair cuts.

There are also correlated pairs that are from the same \pp interaction, 
these are two tracks that share the same ancestry. These pairs can 
arise if there are two \ee pairs in the event from the same parent 
meson, e.g. from a double Dalitz decay of $\pi^0$ or $\eta^0$ or from a 
$\gamma\gamma$ decay where both photons convert in the detector 
material. In this case, the cross-combination of electrons that do not 
result from the same real or virtual photon are possible. Another 
source of these correlated pairs are hadrons from jet fragmentation, 
either within the same jet or in back-to-back jets, that decay into 
electron pairs. These pairs are part of the statistical subtraction 
discussed in the next section.

\subsection{\ee pair spectrum}\label{sec:ee_spec}

Because the source of any electron or positron is unknown, we combine 
all the electrons and positrons in a given event into like-sign 
(\Nlike) foreground pairs, which is defined as sum of pairs of 
electrons and pairs of positrons, and \ee pairs referred to as 
unlike-sign (\Nunlike) foreground pairs. The unlike-sign foreground 
spectrum \Nunlike measures the sum of signal, combinatorial and 
correlated background. For this analysis we use the like-sign pairs to 
determine the backgrounds. The like-sign subtraction method compared to 
the event-mixing technique has the advantage that it also accounts for 
the the correlated pair background that exists in the unlike-sign 
pairs. However, one first needs to correct the like-sign spectrum for 
the relative acceptance difference between \Nlike and \Nunlike pairs.

\begin{figure}[h]
\includegraphics[width=1.0\linewidth]{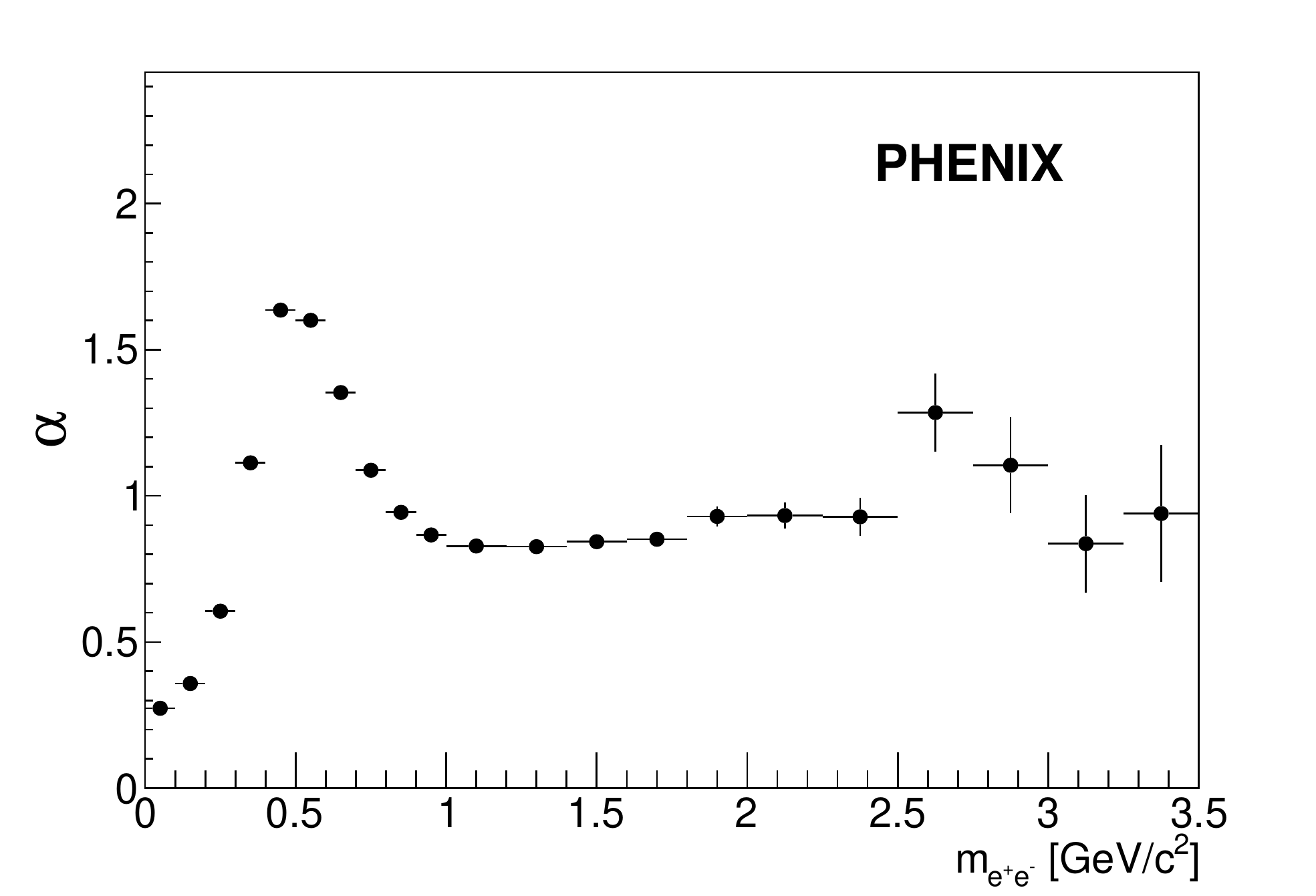}
\caption{\label{fig:fig_alpha}
Relative acceptance correction $\alpha$ defined as the ratio of 
\Bunlike to \Blike as defined in Eq.~\ref{eq_alpha}. This correction 
approaches one at high mass.
}
\end{figure}

The relative acceptance correction $\alpha$ which is purely due to the 
detector geometry is determined via an event mixing technique and is 
given as the ratio of unlike-sign (\Bunlike) to like-sign (\Blike) pair 
spectrum from the mixed events.  The mixed events are generated from 
MB events and are subject to the same requirement as the ERT 
data, i.e. each pair is required to have at least one track above 500 
(or 700) MeV and this track should have fired the ERT trigger. $\alpha$ 
is given by the following equation:

\begin{eqnarray}
\alpha(m,p_T)&=&\frac{\Bunlike(m,p_T)}{\Blike(m,p_T)}\label{eq_alpha}
\end{eqnarray}

\begin{figure}[h]
\includegraphics[width=1.0\linewidth]{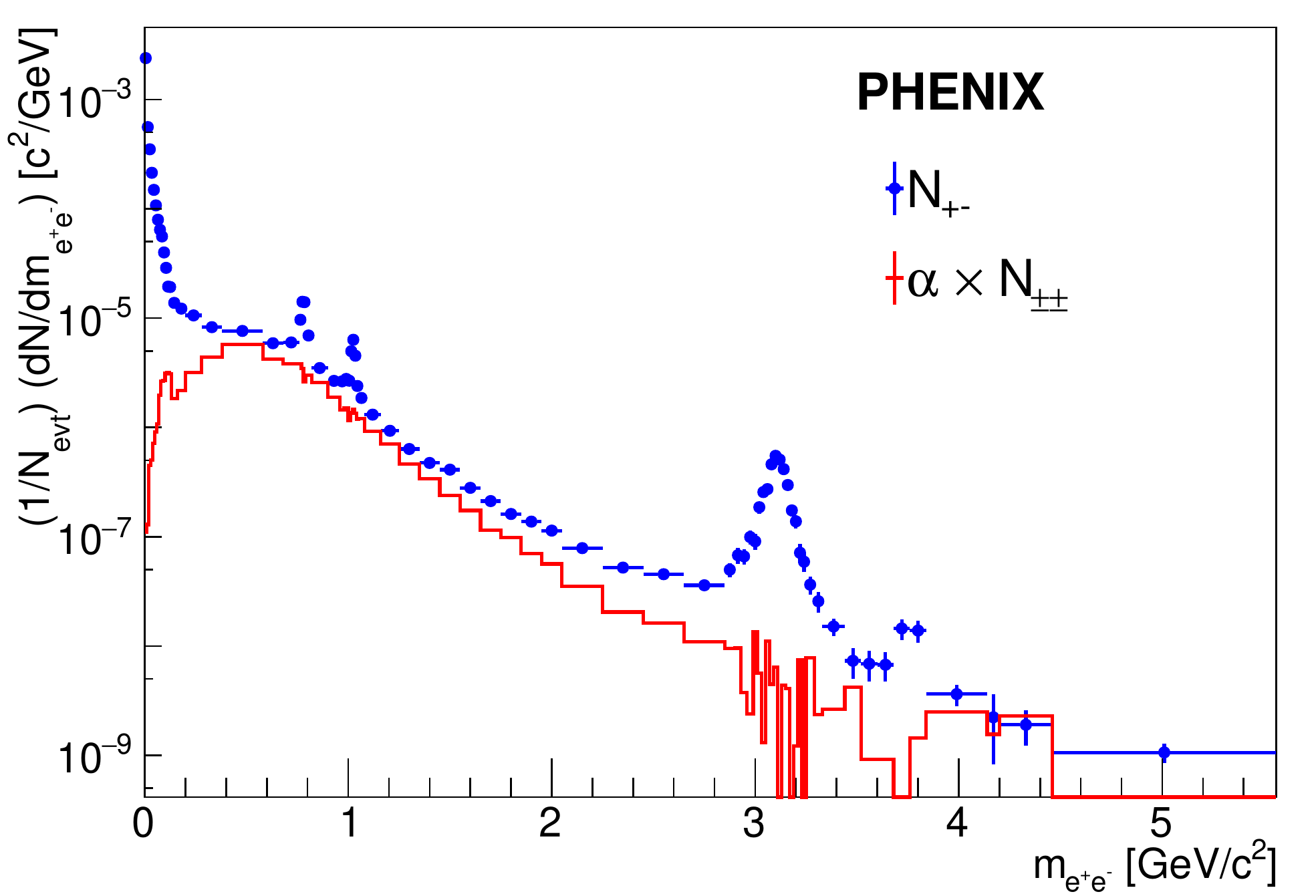}
\caption{\label{fig:signal_bkg}
Unlike-sign foreground \Nunlike spectrum overlaid with like-sign
foreground spectrum after corrected by the relative acceptance
correction $\alpha$ (See Fig.~\ref{fig:fig_alpha} and
Eq.~\ref{eq_alpha}).  
}
\end{figure}

Fig.~\ref{fig:fig_alpha} shows the \pt-integrated $\alpha$ correction 
as a function of mass. The acceptance difference is largest around 
500~MeV/$c^2$. For larger masses, the acceptance difference becomes 
smaller, and consequently $\alpha$ approaches unity as the mass 
increases. In the analysis we apply the $\alpha$-correction double 
differentially in mass and \pt. The errors on the
  $\alpha$-correction are propagated to the final spectrum. For
  systematics, the analysis was checked for \pt dependent fixed
  $\alpha$-values at high masses and results obtained were consistent
  within 5\%. Fig.~\ref{fig:signal_bkg} shows the \pt 
integrated \Nunlike and relative acceptance corrected like-sign mass 
spectrum ($\alpha~ \times$ \Nlike). The acceptance corrected like-sign 
spectrum is subtracted from unlike-sign \Nunlike spectrum to extract 
the signal spectrum, \signal, as defined by Eq.~\ref{eq_signal}.

\begin{equation}  
\signal(m,p_T) = \Nunlike(m,p_T) - \alpha(m,p_T)  
\times \Nlike(m,p_T) \label{eq_signal}
\end{equation}

\subsection{Efficiency corrections}

Eq.~\ref{eq_inv_yld} gives the invariant yield corresponding to a 
\signal pair with mass $m$ and transverse momentum \pt into the PHENIX 
aperture:

\begin{eqnarray}\nonumber\label{eq_inv_yld}
\frac{d^2N_{\ee}}{dm \ dp_{T}} 
&=& \frac{\varepsilon_{\rm BBC}}{N_{\rm BBC}^{\rm sampled}} 
\cdot \frac{1}{\Delta m}
\cdot \frac{1}{\Delta p_{T}}\\\nonumber
&&\cdot \frac{1}{\varepsilon_{\rm rec}(m,p_T)}
\cdot \frac{1} {\varepsilon_{\rm ERT} (m, p_T)}\\
&& \cdot \frac {\signal(m,p_T)}{\varepsilon_{\rm bias}}
\end{eqnarray}

Here $\Delta m$ and $\Delta \pt$ are the bin width in mass and \pt, 
respectively. There are two efficiency corrections that are applied in 
order to obtain the invariant \ee yield. These are the inverse of the 
pair reconstruction efficiency $\varepsilon_{\rm rec}(m,p_T)$ and pair 
trigger efficiency $\varepsilon_{\rm ERT}(m,p_T)$. The 
$\varepsilon_{\rm rec}(m,p_T)$ accounts for losses due to track 
reconstruction, electron identification, pair cuts and detector dead 
areas. The $\varepsilon_{\rm ERT}(m,p_T)$ describes the bias introduced 
by the trigger requirements. Here the BBC efficiency of 
$\varepsilon_{\rm BBC}=0.545~\pm~0.06$ is the fraction of inelastic \pp 
collisions recorded by the BBC. The BBC trigger bias $\varepsilon_{\rm 
bias}$ factor takes into account the fact that for the events with 
tracks in the central arms, the BBC trigger requirement is fulfilled 
only by $0.79~\pm~0.02$ of the events.

The pair reconstruction efficiency $\varepsilon_{\rm rec}(m,p_T)$, as 
well as pair trigger efficiency $\varepsilon_{\rm ERT}(m,p_T)$ are 
determined using a GEANT based simulation of the PHENIX detector. The 
GEANT simulation is tuned to describe the performance of each detector 
subsystem and includes all necessary detector characteristics (dead and 
hot channel maps, gains, noise, etc.).

We simulate \ee pairs with a constant yield in $m,~p_T,~\phi,~|y|<1$, 
and in the mass range $0<m_{\ee}<16$ GeV/$c$$^2$ with \pt in the range 
from 0 to 10 GeV/$c$. These simulated pairs are processed through the 
PHENIX GEANT framework, and are then weighted with the expected yield 
from hadron decays for a given pair $[m,p_T]$. A detailed description 
about pair efficiency and trigger efficiency determination can be found 
in~\cite{Adare:2014iwg,Adare:2009qk,Adare:2008ac}. The efficiency 
corrections are applied double differentially in mass and \pt, and 
similar to the previously published PHENIX dielectron analyses, the 
data are presented in the PHENIX acceptance. The mass spectrum with all 
corrections is shown in Figure~\ref{Fig:pp_mass_spectrum}, together 
with the expected sources discussed in the next section.

\section{Expected pair sources}

The expected yield of \ee pairs from various sources needs to be 
simulated in order to interpret the experimental data. This so called 
cocktail of sources includes the contributions from pseudoscalar and 
vector meson decays, semileptonic decay of heavy flavor, and \ee pairs 
originated via Drell-Yan mechanism.

\subsection{Hadron decays to \ee pairs}

To model the yield of the pseudoscalar mesons $\pi^0$, $\eta$, 
$\eta^{\prime}$, and vector mesons, $\rho$, $\omega$, $\phi$, $J/\psi$, 
$\psi^{\prime}$, $\Upsilon$, we use a detailed fast Monte Carlo 
software package called EXODUS developed within the PHENIX 
framework~\cite{Adare:2009qk}. EXODUS is a phenomenological event 
generator that simulates the particle distributions and their decays. 
EXODUS applies the branching ratios~\cite{Yao:2006px} and decay 
kinematics according to~\cite{Adare:2006hc}. External bremsstrahlung in 
the PHENIX detector material is approximated by placing all the 
detector material to be traversed by the electron at the radius of the 
beampipe. The pair mass distribution from Dalitz decays 
($\pi^0,\eta,\eta^{\prime} \rightarrow ee\gamma$) and $\omega 
\rightarrow ee\pi^0$ follows the Kroll-Wada 
expression~\cite{Kroll:1955zu} multiplied by the electromagnetic form 
factors measured by the Lepton-G 
collaboration~\cite{Landsberg:1986fd,Dzhelyadin:1980tj}.  The vector 
mesons ($\rho,\omega,\phi,J/\psi,\psi^{\prime}\rightarrow\ee)$ are 
assumed to be unpolarized and for their decay the Gounaris/Sakurai 
expression is used~\cite{Gounaris:1968mw}. For the Dalitz decays in 
which the third body is a photon, the angular distribution is sampled 
according to $1+\lambda cos^{2}\theta_{CS}$ distribution. $\theta_{CS}$ 
is the polar angle of the electrons in the Collins-Soper frame and 
$\lambda$ is an angular parameter.

The hadrons are generated with a uniform rapidity density \dndy within 
$|\eta|\leq$0.35 and a homogeneous azimuthal distribution in 2$\pi$. 
Once generated, these hadrons are filtered through the ideal PHENIX 
acceptance while applying the measured momentum resolution from the 
data. The key input is the parameterization of the \pt dependence of 
the invariant cross section of neutral pions. To obtain this reference 
we fit the \pt distribution of $\pi^0$ and $\pi^{\pm}$ data, as 
reported by PHENIX~\cite{Adler:2003pb,Adare:2007dg,Adare:2011vy}, to a 
modified Hagedorn function (Eq.~\ref{eq_mod_hag}):

\begin{equation}\label{eq_mod_hag}
  E\frac{d\sigma^3}{dp^3} = A(e^{-(ap_T+bp_T^2)} + p_T/p_0)^{-n}
\end{equation}

The fit parameters and resulting \dndy values for \pp collisions are 
tabulated in Table~\ref{tab:fit_pars}. These values supersede 
those published in~\cite{Adare:2009qk,Adare:2008ac} as they are based 
on new and/or more precise data from larger data sets. The pion 
parameterization determined here deviates by about 3\% from the one 
used in earlier publications.

\begin{table}[h]
 \caption{Fit parameters for \pp collisions derived from a simultaneous fit to
  the $\pi^0$ and charged pions data using the modified Hagedorn 
  function~(Eq.~\ref{eq_mod_hag}).}
  \label{tab:fit_pars}
\begin{ruledtabular}\begin{tabular}{cc}
     Parameter & Value \\
     \hline
     \dndy   & 1.139 $\pm$ 0.10\\
     $A$[$mb GeV^{-2}c^{3}$]  &  492 $\pm$ 67  \\
     $a$[$(GeV/$c$)^-1$]       &  0.266 $\pm$ 0.031\\
     $b$[$(GeV/$c$)^-2$]       &  0.092 $\pm$ 0.021\\
     $p_0$[$GeV/$c$$]        & 0.68 $\pm$ 0.02 \\
     $n$                     &  8.27 $\pm$ 0.07\\
  \end{tabular} \end{ruledtabular}
\end{table}

\begin{figure}[htb]
  \vspace{-3mm}
  \begin{center}
    \includegraphics[width=1.0\linewidth]{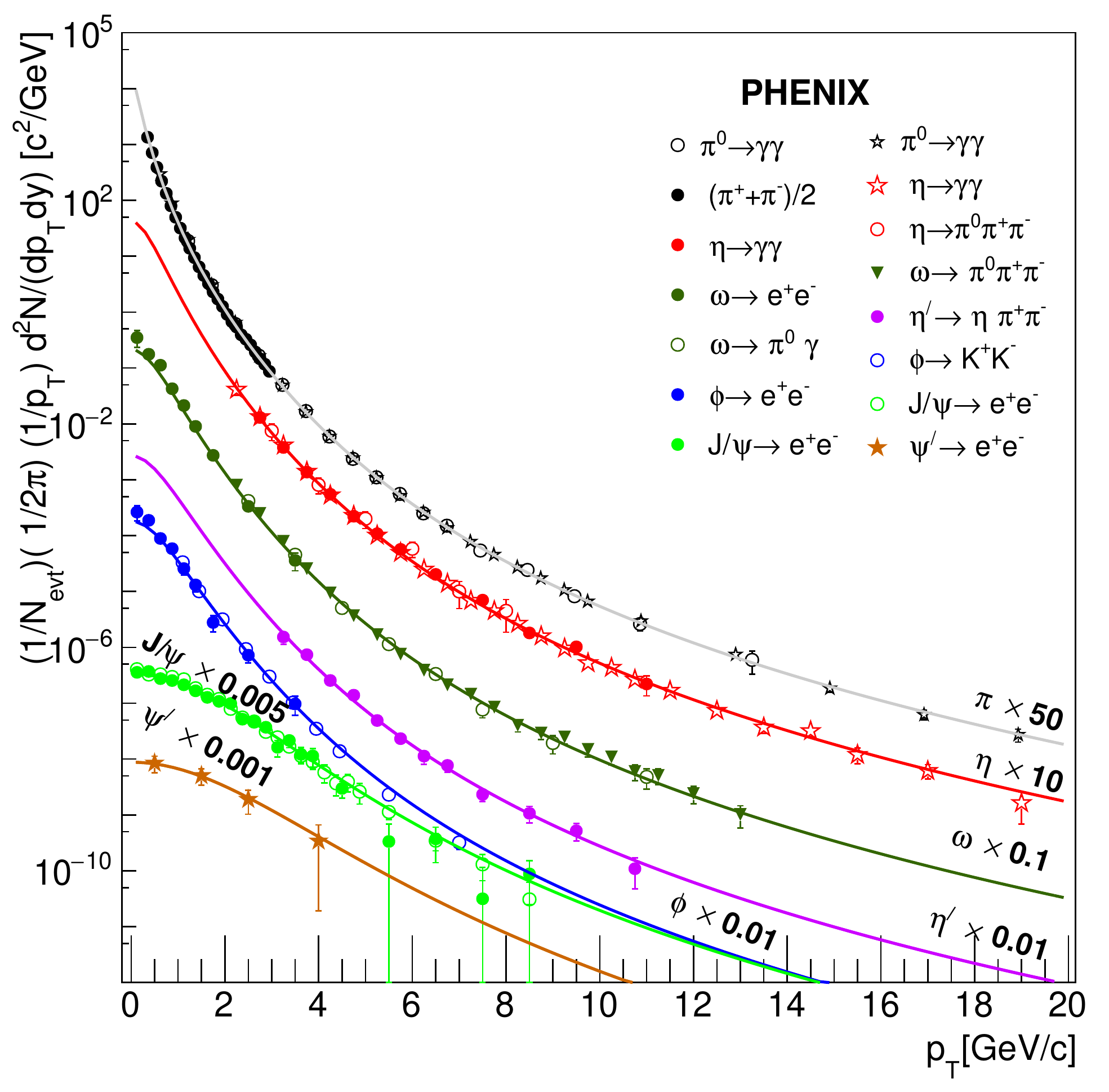}
  \end{center}
  \vspace{-3mm}
  \caption{Compilation of meson production in \pp collisions at 
  \sqsn~=~200 GeV. The data shown above are taken from the following
  sources:
  $\pi^0\rightarrow\gamma\gamma$~\cite{Adler:2003pb,Adare:2007dg},
  $(\pi^+ +\pi^-)/2$~\cite{Adare:2011vy},
  $\eta\rightarrow\gamma\gamma$~\cite{Adare:2010cy, Adler:2006bv},
  $\eta\rightarrow\pi^0\pi^+\pi^-$~\cite{Adler:2006bv},
  $\omega\rightarrow e^+e^-$~\cite{Adare:2010fe},
  $\omega\rightarrow\pi^0\pi^+\pi^-$~\cite{Adare:2010fe},
  $\omega\rightarrow\pi^0\gamma$~\cite{Adare:2010fe}, $\phi\rightarrow
  K^+K^-$~\cite{Adare:2010pt}, $\phi\rightarrow
  e^+e^-$~\cite{Adare:2010fe}, J$\psi\rightarrow
  e^+e^-$~\cite{Adare:2006kf,Adare:2011vq}, $\psi^{\prime}\rightarrow
  e^+e^-$~\cite{Adare:2011vq}.
  The data are compared to the parameterization based on
  $m_T$ scaling used in EXODUS.}
  \label{fig:fig_mesons}
\end{figure}

The \pt distribution of other mesons is parameterized by fixing all but 
the normalization parameter ($A$) from the pion spectrum, and assuming 
scaling with \mt, i.e. replacing \pt by $\sqrt{(\pt^2 -(m_{\pi0}c)^2 + 
(m_{h}c)^2)}$, where $m_h$ is the mass of the hadron. The normalization 
parameter $A$ relates the total \dndy of a given hadron to the \dndy of 
the pions. The successful description of \mt scaling is apparent in 
Fig.~\ref{fig:fig_mesons} which shows measured \pt spectra of various 
mesons as published by PHENIX. In order to extract the meson yield the 
fits were integrated over all the \pt. For the $\rho$ meson, we assume 
$\sigma_{\rho}/\sigma_{\omega}= 1.15~\pm~0.15$ consistent with the 
values found in the jet fragmentation~\cite{Yao:2006px}.

\begin{table}[hbt]
 \caption{Rapidity density for the mesons extracted from the fits and
  used in the EXODUS decay generator.}
  \label{tab:dndy_mesons}
  \begin{ruledtabular}\begin{tabular}{cccc}
    Meson   & $dN/dy|_{y=0}$ & Data source\\
    \hline
    $\pi^0,~\pi^+,\pi^-$    & 1.139 $\pm$ 0.10& \cite{Adler:2003pb,Adare:2007dg,Adare:2011vy}\\
    $\eta$         & 0.093 $\pm$ 0.0002   & \cite{Adler:2006bv,Adare:2010cy}\\
    $\omega$       & 0.0744 $\pm$ 0.0017  & \cite{Adare:2010fe}\\
    $\phi$         & 0.009  $\pm$ 0.0002  &\cite{Adare:2010pt,Adare:2010fe}\\
    $\eta^{\prime}$ & 0.0123 $\pm$ 0.0008 & \cite{Adare:2010fe}\\
    $J/\psi$       & 1.74$\times10^{-05}~\pm$ 5.1$\times10^{-7}$& \cite{Adare:2006kf,Adare:2011vq}\\
    $\psi^{\prime}$ & 3.1$\times10^{-06}~\pm$ 6.2$\times10^{-7}$& \cite{Adare:2011vq}\\
  
  \end{tabular}\end{ruledtabular}
\end{table}

A compilation of the \dndy values for the various mesons extracted from 
the fits and the references for the data used are shown in 
Table~\ref{tab:dndy_mesons}. These values agree with those from 
\cite{Adare:2009qk,Adare:2008ac} within the systematic uncertainties. 
The differences reflect that more precise data for the pion and other 
mesons are available today.

\subsection{\ee pairs from Drell Yan}

We used \pythia event generator with same settings as mentioned in 
\cite{Adare:2014iwg} to simulate \ee pairs from the Drell-Yan 
mechanism. For the normalization we used a cross section of 42 nb as 
was used in \cite{Adare:2014iwg,Adare:2008ac}. We also performed a 
study where the DY contribution was left as a free parameter. This 
affected the \bb cross section by 20\% and we assigned that as a 
systematic uncertainty on the cross section determination.

\subsection{Heavy flavor contribution to \ee pairs}

The \ee pairs that originate from the semileptonic decays of \cc and 
\bb are collectively referred to as heavy flavor pairs. The heavy 
flavor yield was simulated using three different event generators. The 
details of these event generators are described below. 

\subsubsection{\pythia} 

\pythia~\cite{Sjostrand:2006za} is a multi-purpose leading order event 
generator. It generates heavy quark pairs with massive matrix elements 
and fragmentation and hadronization is based on the Lund string model. 
Additional transverse momentum is generated in \pythia by virtue of the 
assumed intrinsic (primordial) transverse momentum $k_T$. We used 
\pythia in forced \cc or \bb production mode, and CTEQ5L was used as 
the input parton distribution function. The same settings as published 
in the \dA paper~\cite{Adare:2014iwg} are hereby used.

\subsubsection{\mcnlo}

The \mcnlo (Monte Carlo at next-to-leading order) formalism is described 
in detail in~\cite{Frixione:2002ik,Frixione:2003ei}, and is a method 
for matching next-to-leading order (NLO) QCD calculations to parton 
shower Monte Carlo (pSMC) simulations.  Parton showers will generate 
terms that are already present in the NLO calculations. To avoid double 
counting, the \mcnlo scheme removes such terms from the NLO expression. 
As a result, \mcnlo output contains events with negative weight.

In this work, \mcnlo v4.10 (interfaced with {\sc herwig}v{\footnotesize 
6.521}~\cite{Corcella:2000bw}) was used. The default package was 
altered to enable charm production by changing the process code from 
-1705 ($H_1H_2 \rightarrow \bb + X$) to -1704 ($H_1H_2 \rightarrow \cc 
+ X$) and the heavy quark mass was adjusted to the charm quark mass 
i.e. 1.29 GeV/$c^2$. $H_{1,2}$ represent hadrons (in practice, nucleons 
or antinucleons). The bottom quark mass was set to 4.1 GeV/$c^2$. The 
default scale choice was used:

\begin{equation} 
\mu_0^2=\frac{1}{2}(m_{T}^2(Q)+m_{T}^2(\bar{Q})),
\end{equation}

\noindent where $m_{T}^2=p_{T}^{2}+m^2$ and \pt is the transverse 
momentum of the heavy flavor in the underlying Born configuration. $Q$ 
and $\bar{Q}$ correspond to the heavy quark and antiquark. No other 
parameters were modified. CTEQ6M~\cite{Pumplin:2002vw} was used to 
provide the input parton-distribution function.

\begin{figure*}[htb]
\includegraphics[width=0.998\linewidth]{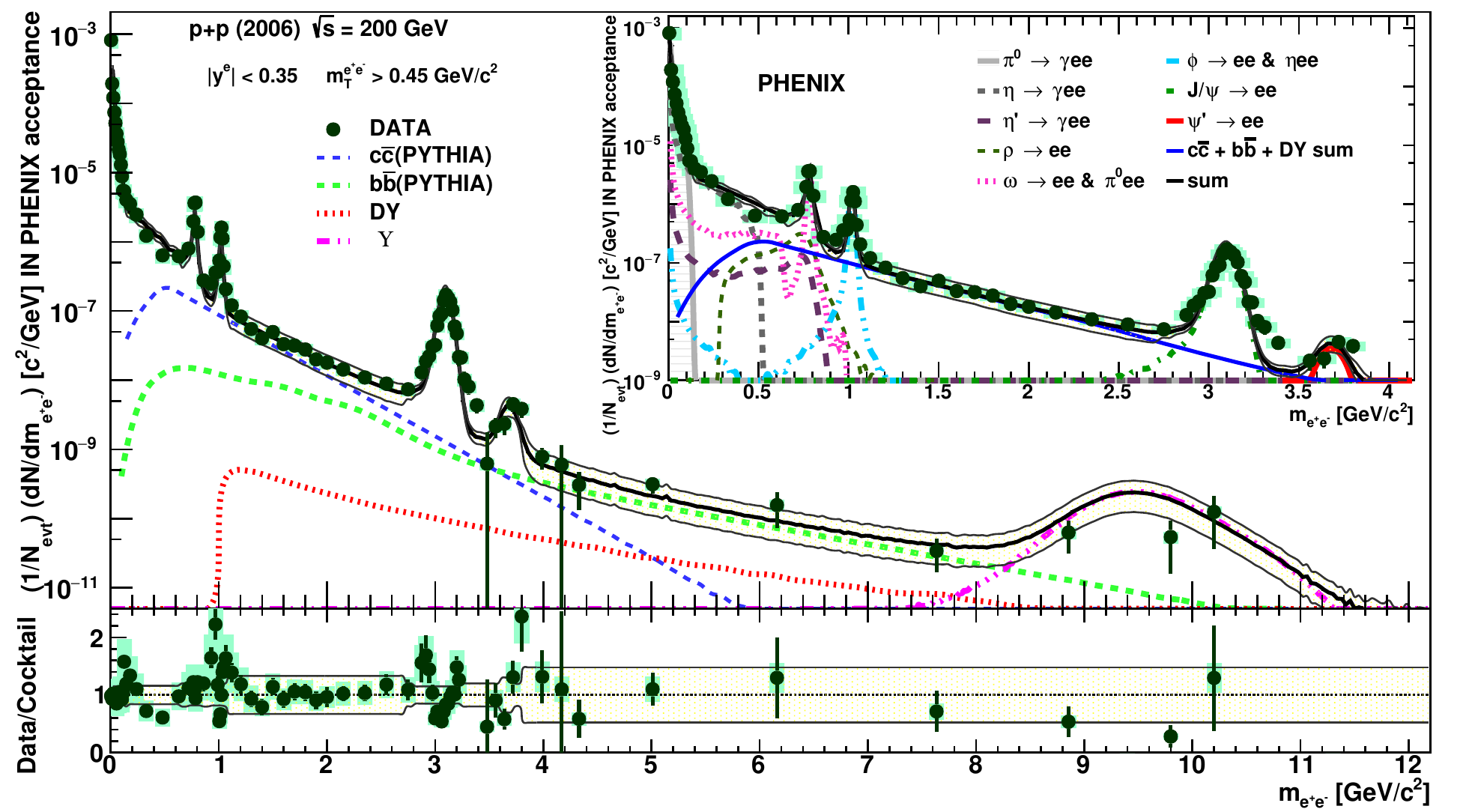}
\caption{\label{Fig:pp_mass_spectrum}
Inclusive \ee pair yield from \pp collisions as a 
function of mass. The data are compared to our model of expected 
sources. The inset shows in detail the mass range up to 4.5 
GeV/$c^2$. In the lower panel, the ratio of data to expected 
sources is shown with systematic uncertainties.
} 
\end{figure*}
\begin{figure*}[htb]
\begin{minipage}{0.75\linewidth}
  \includegraphics[width=1.0\linewidth]{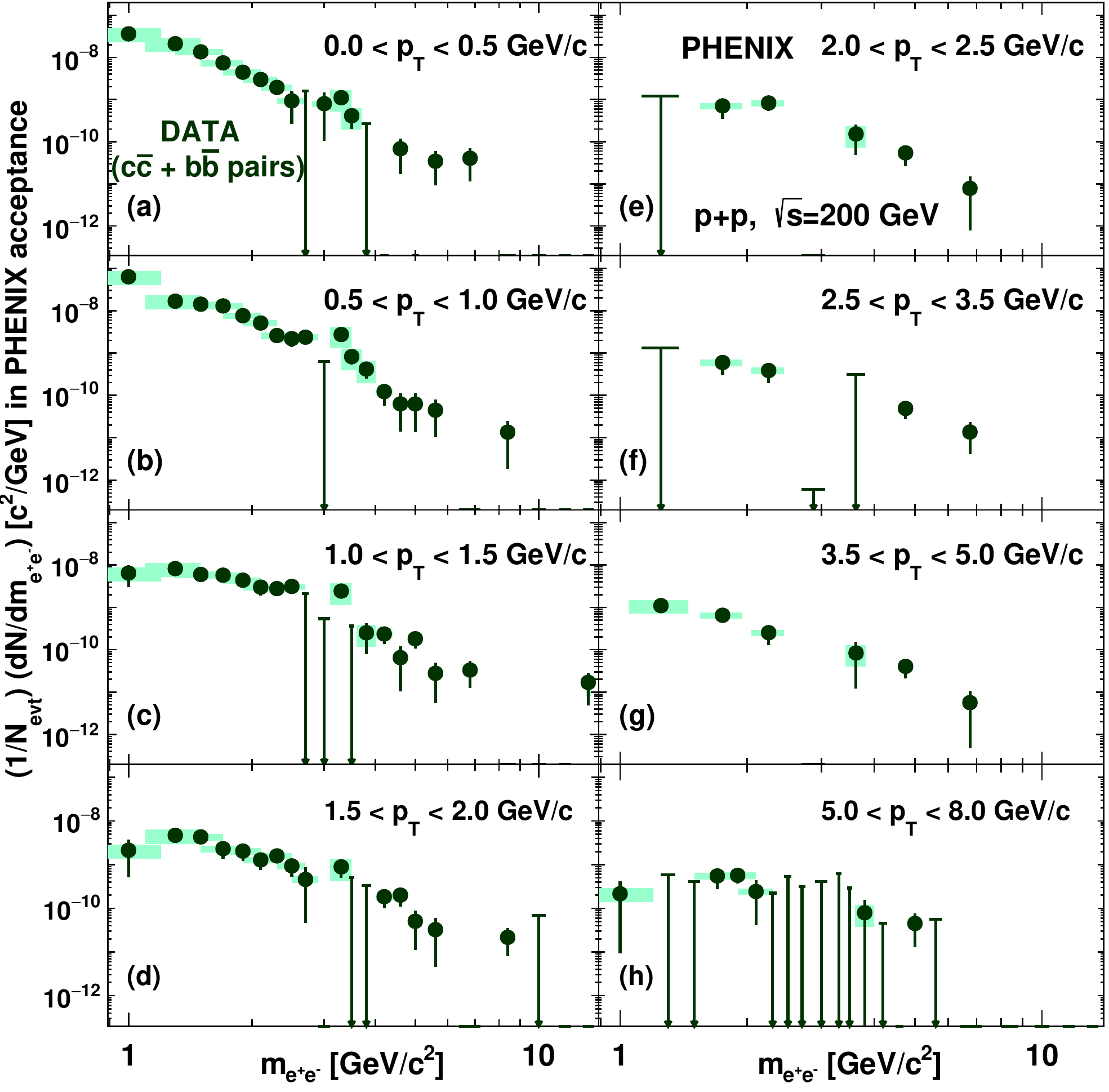}
\end{minipage}
\hspace{0.2cm}
\begin{minipage}{0.15\linewidth}
  \caption{\label{Fig:doublediffdata}
Double differential \ee pair yield from semi-leptonic decays of heavy 
flavor in inelastic \pp collisions. Shown are mass projections in 
slices of \pt. The \pt intervals are indicated in each panel. 
Systematic uncertainties are shown as boxes, downward pointing arrows 
indicate upper limits at 90\% CL.
}
\end{minipage}
\end{figure*}

 \subsubsection{\powheg}
The \powheg (Positive Weight Hardest Emission Generator) formalism is 
described in detail in~\cite{Frixione:2007nw}. Compared to \mcnlo, 
\powheg generates positive weighted events only, and can be interfaced 
to any shower MC that is either \pt-ordered (e.g. \pythia), or allows 
the implementation of a \pt veto (e.g. \herwig++), while avoiding any 
double counting when matching NLO calculations and parton shower Monte 
Carlo. In this work, \powheg v1.0 was interfaced with \pythia 
v8.100~\cite{Sjostrand:2007gs}. Parton showering in \pythia is \pt 
ordered and merges naturally with \powheg. CTEQ6M~\cite{Pumplin:2002vw} 
was used to provide the input parton distribution function. Similar to 
the other two frameworks, the charm and bottom masses were set to 1.29 
GeV/$c^2$ and 4.1 GeV/$c^2$ respectively. The default scale choice was 
used:

\begin{equation} 
\mu_0^2=p_{T}^{2}+m^2,
\end{equation}

\noindent where \pt is the transverse momentum of the heavy flavor in 
the underlying Born configuration. No other parameters were modified.

The electrons and positrons from all the above mentioned generators are 
filtered through the PHENIX acceptance~\cite{Adare:2009qk} and are 
folded with the experimental momentum resolution as well as with the 
energy loss due to bremsstrahlung. The \ee pair acceptance depends on 
the production process, which determines the correlation between the 
electron and positron. More detailed description about the \ee pair 
acceptance on (i) the QCD production of the \qq pair and (ii) the decay 
kinematics of the two independent semi-leptonic decays has been 
discussed in~\cite{Adare:2014iwg}. Because the heavy flavor generators 
discussed above treat the \qq correlations differently, the number of 
\ee pairs that fall into PHENIX acceptance varies from one generator to 
the other.

\section{Systematic uncertainties}

In this section we summarize the systematic uncertainties on the data 
and expected sources. Systematic uncertainties on the data are due to 
limitations in the determination of the relative acceptance correction, 
the electron identification efficiency, model input used to evaluate 
the efficiency, and the ERT trigger efficiency. These uncertainties are 
evaluated by varying all the electron identification cuts and pair 
cuts, by varying the ERT trigger efficiency within its statistical 
accuracy and by using different cuts and sub-samples of the data to 
determine the relative acceptance correction. For all the variations 
the final result was determined and found stable within the quoted 
systematic uncertainties.

The main systematic uncertainties on the hadron cocktail comes from the 
measured uncertainty on the \dndy of pions. For the heavy flavor part 
of the cocktail, the assigned uncertainty to \cc and \bb normalization 
comes from this analysis.

Table~\ref{tbl:tbl_sys_all} gives a summary of the systematic errors. 
The total systematic error on data are added in quadrature and the same 
is done for the expected sources.

\begin{table}[htb]
\caption{Summary of the various systematic uncertainties considered in this analysis.}
\label{tbl:tbl_sys_all}
\centering
\small\addtolength{\tabcolsep}{-3pt}
\begin{ruledtabular}\begin{tabular}{ccc}
    
    Source & \multicolumn{2}{c}{Syst. uncertainty} \\
           & (mass $\leq1.0$GeV/$c^{2}$)&  (mass $>1.0$ GeV/$c^{2}$) \\
\\
    \multicolumn{3}{c}{Data systematics}\\
    eID       & 15\%    & 10\%  \\
    Input model & 15\%    & 15\%  \\
    ERT       & 10\%    & 5\%  \\
    Fiducials & \multicolumn{2}{c}{10\%} \\
    $\alpha-$ correction & \multicolumn{2}{c}{5\%} \\
    BBC bias & \multicolumn{2}{c}{10\%} \\
\\
    \multicolumn{3}{c}{Cocktail systematics}\\  
    Hadronic cocktail    & \multicolumn{2}{c}{20\%} \\ 
    \cc cross section    & \multicolumn{2}{c}{32\%}\\
    \bb cross section    & \multicolumn{2}{c}{36\%}\\
\end{tabular} \end{ruledtabular}
\end{table}
 
\section{Results}
\subsection{Heavy-flavor \ee pairs from \pp collisions}
\label{results_pp} 

Figure~\ref{Fig:pp_mass_spectrum} shows the measured double 
differential \ee pair yield in the PHENIX acceptance projected onto the 
mass axis. The figure also shows the distributions of \ee pairs from 
charm, bottom and Drell-Yan obtained using the \pythia event generator. 
The mass region below 1.0 GeV/$c$$^2$ is comprised of resonances and a 
continuum dominated by three body decays of pseudoscalar and vector 
mesons. In this mass region, all cocktail contribution, with exception 
of the heavy flavor meson decay contributions, are absolutely 
normalized as discussed previously. The contributions of various 
hadronic decay sources to the cocktail are shown in the inset that 
highlights the mass spectrum up to 4.5 GeV/$c^2$. The mass spectrum 
above 1.0 GeV/$c^2$ is dominated by the \ee pairs from decays of heavy 
flavor mesons. The heavy flavor contributions to the dilepton continuum 
above 1.0 GeV/$c$$^2$ are normalized to the data. Good agreement 
between data and cocktail over the entire mass range is evident from 
the ratio of data to the cocktail shown in the lower panel of 
Fig.~\ref{Fig:pp_mass_spectrum}. We note that below 0.6 MeV/$c^2$ there 
are large systematic uncertainties resulting from the ERT trigger 
efficiency correction. In this mass region, the results published 
in~\cite{Adare:2008ac} are more accurate due to large sample of MB
data available for that analysis. The bulk of the 2006 data used 
here was taken with the ERT trigger. Our current heavy flavor analysis 
is based on the mass region above 1.16 GeV/$c^2$ and thus not affected 
by systematic uncertainties around 0.5 GeV/$c^2$.

The \ee pair spectrum from heavy flavor decays is determined using the 
technique developed for \dA collisions~\cite{Adare:2014iwg}. The 
expected yield of \ee pairs from pseudoscalar and vector meson decays 
as well as Drell-Yan pairs is subtracted from the \ee pair spectra. The 
subtraction is done double differentially in mass and \pt. The 
resulting mass spectra of \ee pairs from heavy flavor decays are shown 
in Fig.~\ref{Fig:doublediffdata} for different pair \pt ranges. Below 
1.0 GeV/$c$$^2$, the yield of \ee pairs is dominated by hadronic decay 
contributions and after the subtraction the \ee pair yield from heavy 
flavor decays cannot be extracted with sufficient accuracy. Therefore 
Fig.~\ref{Fig:doublediffdata} is truncated just below 1 GeV/$c^2$. For 
those mass regions above 1 GeV/$c^2$ where the inclusive \ee yield is 
dominated by vector meson decays to \ee the subtracted yield can not be 
determined accurately, and hence upper limits are quoted for the 
subtracted spectra. We use \pt bins of width of 500 MeV/$c$ up to \pt = 
2.5 GeV/$c$. For pair $\pt>3.0$ GeV/$c$, statistical limitations 
dictate the use of broader \pt bins.

\begin{figure*}[htb]
\begin{minipage}{0.49\linewidth}
  \includegraphics[width=0.99\textwidth]{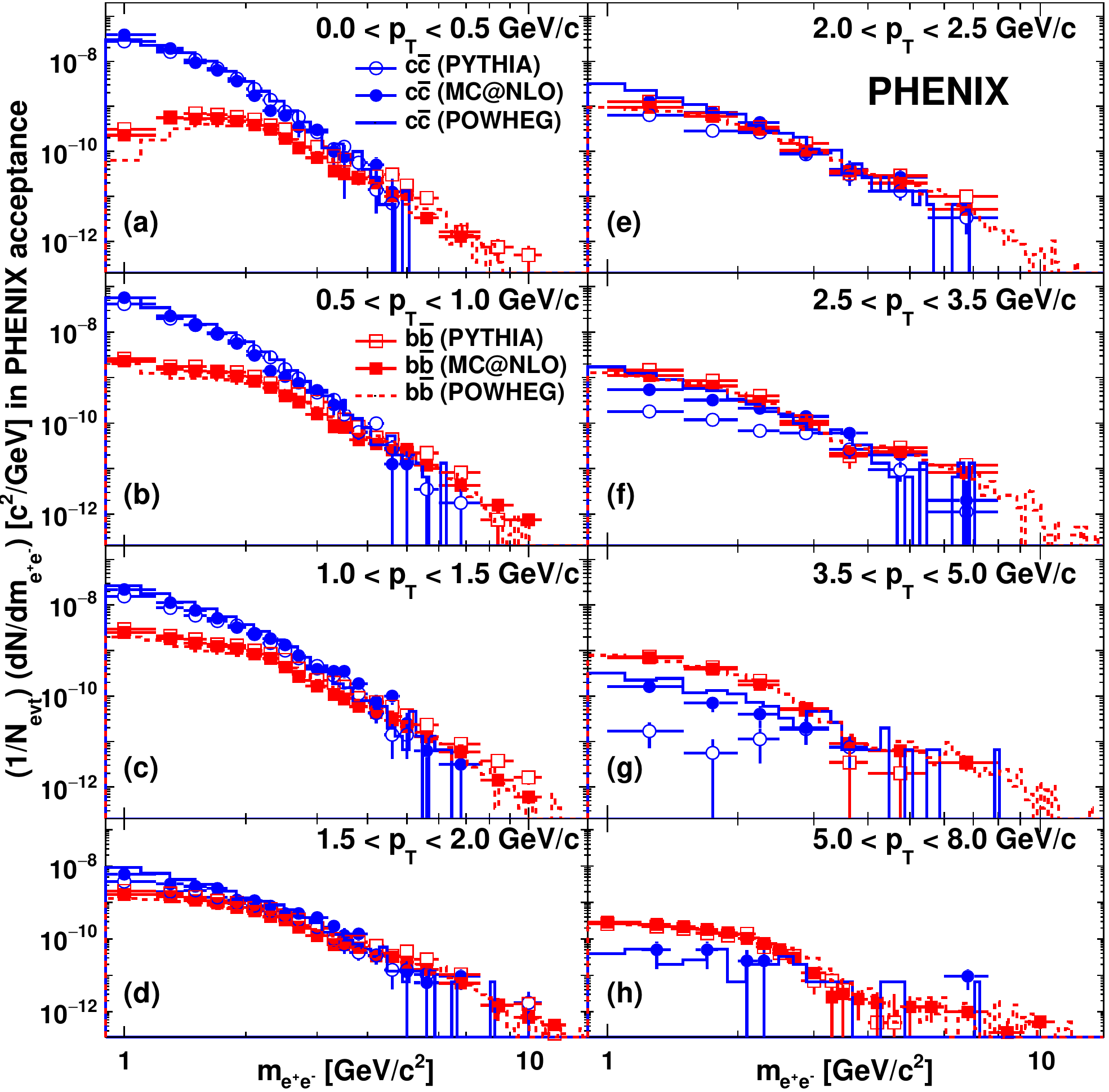}
\caption{Double differential \ee pair yield from semi-leptonic decays 
of heavy flavor as simulated by \pythia, \mcnlo, and \powheg. Shown are 
mass projections in slices of \pt. The \pt intervals are indicated in 
each panel.}
  \label{Fig:doublediffsim}
\end{minipage}\hfill
\begin{minipage}{0.49\linewidth}
\includegraphics[width=0.99\linewidth]{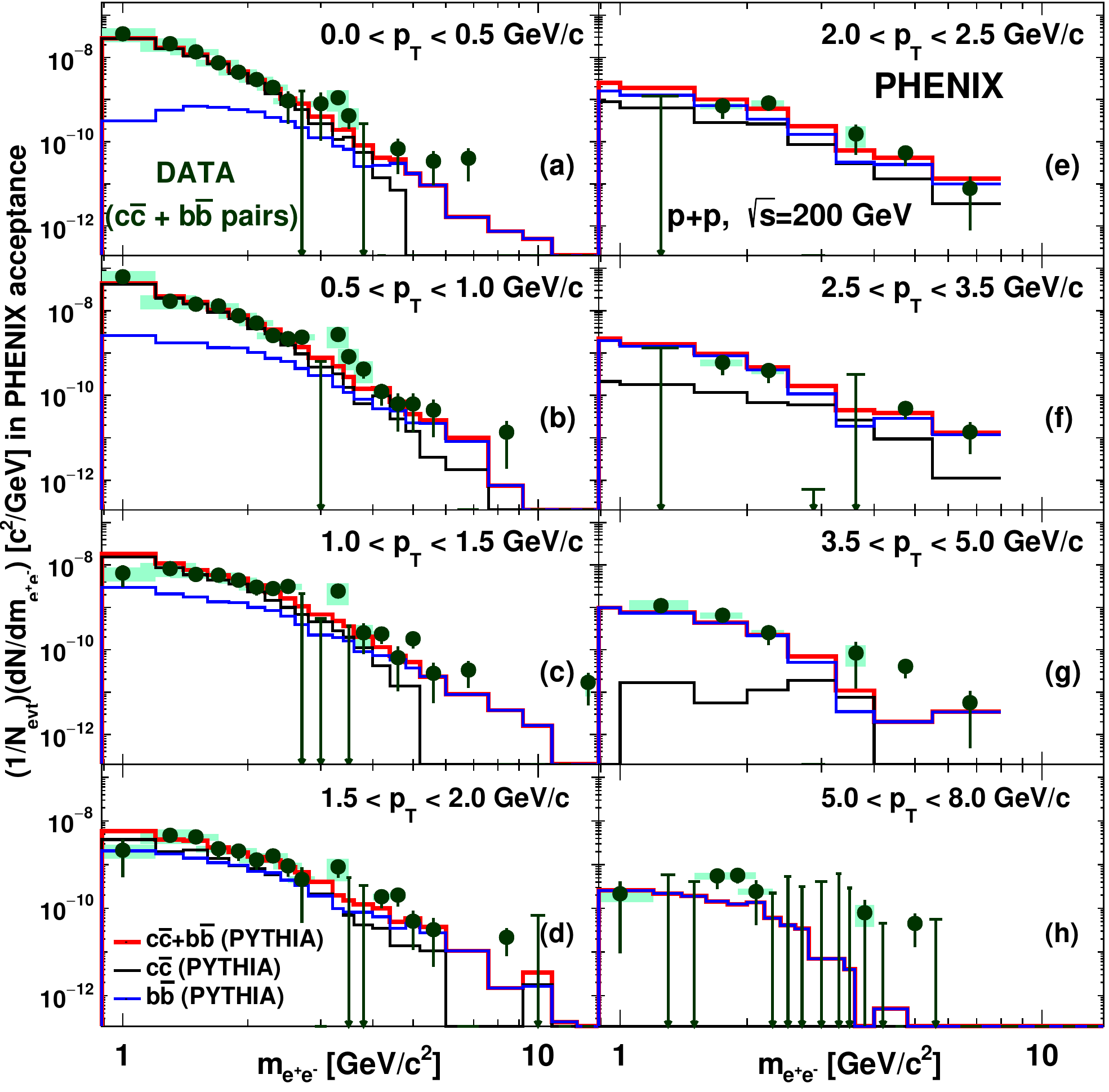}
\caption{\label{Fig:fig_data_pythia} Double differential \ee pair yield 
from heavy-flavor decays fitted to simulated distributions from
\pythia. The simulation is fitted to data in the mass region between
1.15 $<\mee <$ 2.4 GeV/$c^2$ and 4.1 $<\mee <$ 8.0 GeV/$c^2$.}
\end{minipage}\hfill
\begin{minipage}{0.49\linewidth}
\includegraphics[width=0.99\linewidth]{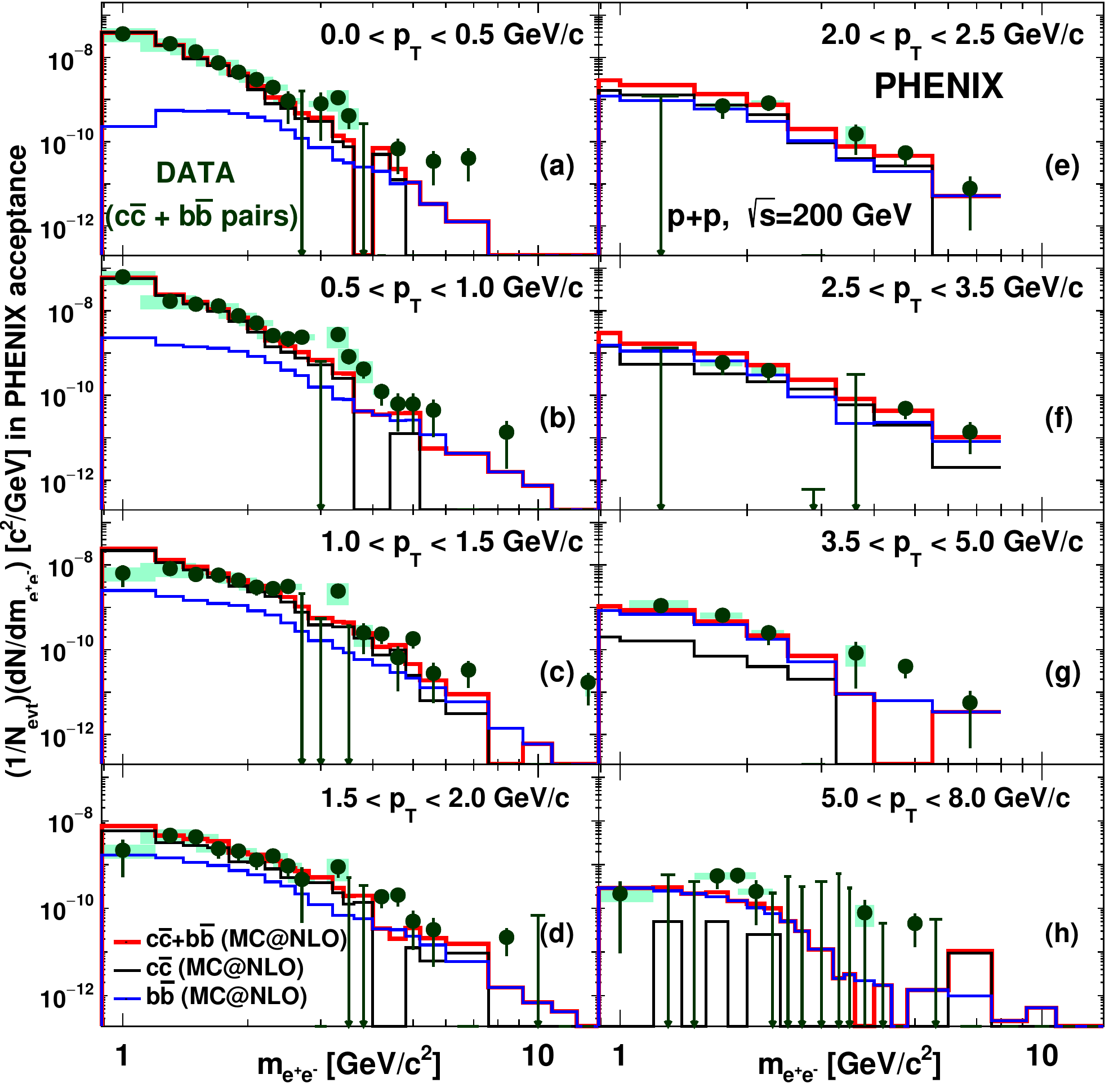}
\caption{\label{Fig:fig_data_mcnlo} Double differential \ee pair yield 
from heavy-flavor decays fitted to simulated distributions from
\mcnlo.\\ The simulation is fitted to data in the mass region between
1.15 $<\mee <$ 2.4 GeV/$c^2$ and 4.1 $<\mee <$ 8.0 GeV/$c^2$.}
\end{minipage}
\begin{minipage}{0.49\linewidth}
\includegraphics[width=0.99\linewidth]{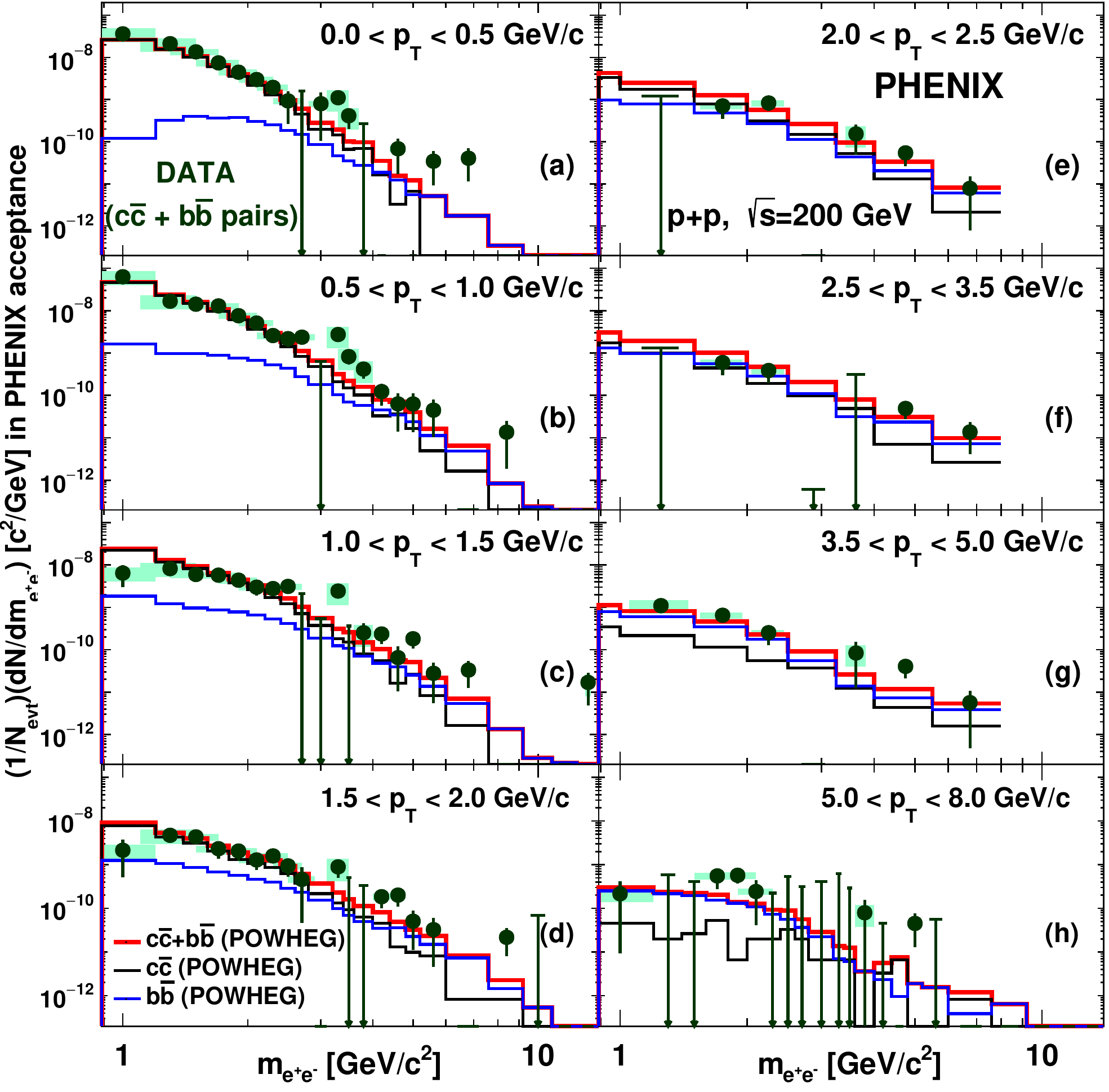}
\caption{\label{Fig:fig_data_powheg} Double differential \ee pair yield 
from heavy-flavor decays fitted to simulated distributions from
\powheg. The simulation is fitted to data in the mass region between
1.15 $<\mee <$ 2.4 GeV/$c^2$ and 4.1 $<\mee <$ 8.0 GeV/$c^2$.}
\end{minipage}\hfill
\end{figure*}

The \ee pair distributions from heavy flavor decays were simulated 
using three Monte Carlo generators, \pythia, \mcnlo, and \powheg with 
parameter settings as discussed above. The results are shown in 
Fig.~\ref{Fig:doublediffsim}. The three generators are compared using 
the normalization

\begin{table*}
\caption{Summary of \cc and \bb cross section measured in \pp collisions 
   using three different generators \pythia, \mcnlo, and \powheg. These are 
   obtained by extrapolating to $4\pi$ the fitting results from the measured
   \ee pairs from heavy flavor.}
    \label{tab:xsec_pp}
\begin{ruledtabular}  	\begin{tabular}{cccc}
     \pp      & \pythia ($\mu$b)    &   \mcnlo ($\mu$b) & \powheg ($\mu$b)\\
      \cc  & 356 $\pm$ 27 (stat) $\pm$ 89(syst) & 708 $\pm$ 55 (stat) $\pm$ 175 (syst) & 267 $\pm$ 19 (stat) $\pm$ 67 (syst)\\
     \bb & 4.81 $\pm$ 0.71 (stat)$\pm$ 1.00 (syst) & 3.85 $\pm$ 0.73 (stat)$\pm$ 0.8 (syst) & 2.91 $\pm$ 0.63 (stat)$\pm$ 0.61 (syst)\\
   \end{tabular}  \end{ruledtabular}
 \end{table*}

\noindent obtained from fitting the data to the respective event
generators as described below. As seen in 
Fig.~\ref{Fig:doublediffsim} and already described in detail 
in~\cite{Adare:2014iwg}, the separation of \ee pairs from \cc and \bb 
is more evident when one simultaneously analyzes mass and \pt of the 
pairs. The yield from \cc is dominant for masses below 3 GeV/$c^2$ and 
pair \pt less than 2 GeV/$c$, whereas \bb is dominant across all mass 
region for higher \pt. For the pairs with \pt $>$ 3.5 GeV/$c$, the 
largest contribution to the \ee yield comes from single $b$ decay 
chains with a semileptonic decay of the parent B meson followed by a 
semileptonic decay of the daughter D meson.

The generated distributions are fitted simultaneously to all data in 
\pt and mass in the mass regions between 1.15 $<\mee <$ 2.4 GeV/$c^2$ 
and 4.1 $<\mee <$ 8.0 GeV/$c^2$. The mass region from 2.4 to 4.15 
GeV/$c^2$ is excluded to avoid any remnant contributions to the \ee 
yield from $J/\psi$ and $\psi'$ decays after the subtraction. Such 
remnant yield could result from an imperfect description of the line 
shapes, in particular of the low mass tail due to bremsstrahlung. For 
each MC generator there are two independent parameters that are fitted, 
which are the \cc and \bb cross sections in $4\pi$. 
Figs.~\ref{Fig:fig_data_pythia},~\ref{Fig:fig_data_mcnlo}, and 
\ref{Fig:fig_data_powheg} show the comparison of fitted distributions 
to the data for \pythia, \mcnlo, and \powheg, respectively.  The 
$\chi^2/NDF$ values are 1.2, 1.5, and 1.4 for \pythia, \mcnlo, and 
\powheg, respectively, with an $NDF$ equal to 65. Here, 
only statistical errors are used in the fit.  Because the \cc simulated 
pairs have smaller statistics at high masses for \pt$>$5 GeV/$c$, we 
include the errors from simulations into the fitting routine.  Any 
improvement from additional statistics is expected to be minimal unless 
significant computing resources are allocated.

\begin{figure*}[!htb]
\includegraphics[width=0.75\linewidth]{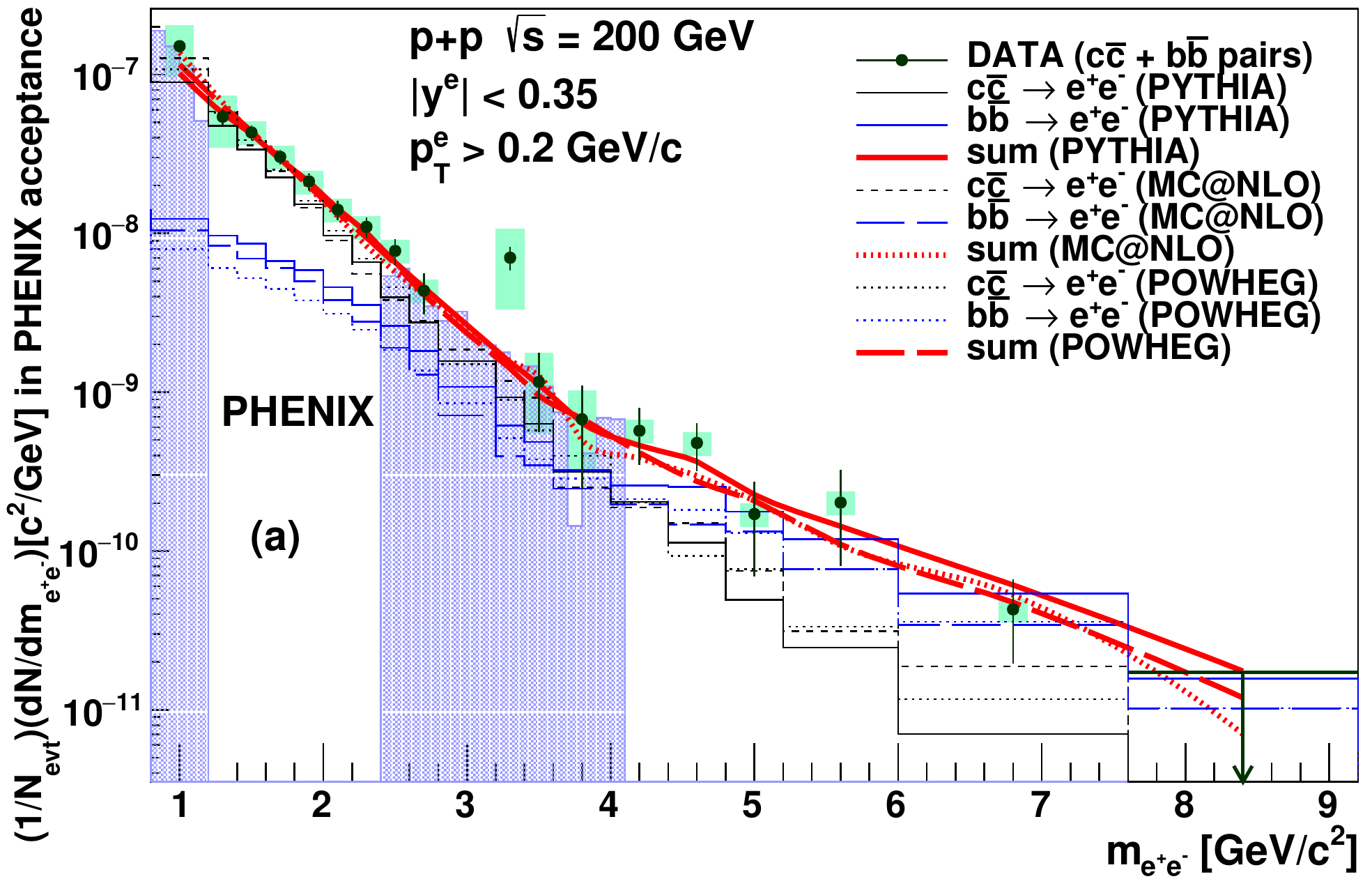}
\includegraphics[width=0.75\linewidth]{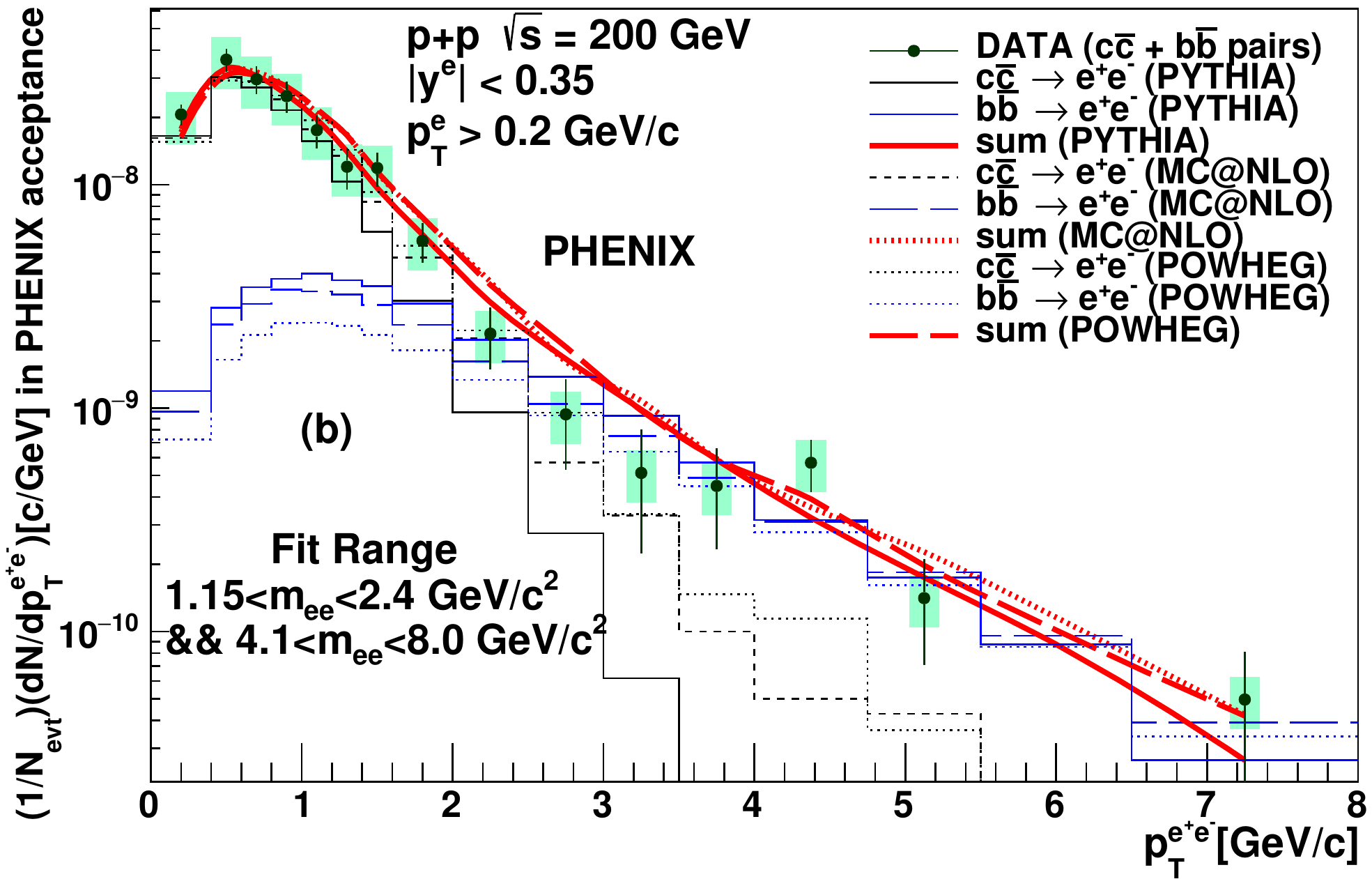}
\caption{\label{Fig:HFprojections}
The top panel compares the mass dependence of \ee pair yield with \pythia, 
\mcnlo and \powheg calculations. The bottom panel shows the comparison 
for the \pt dependence. The blue region shown in the top panel is not used 
in the fitting and is excluded in the \pt projection.
}
\end{figure*}

\begin{figure*}[!htb]
\includegraphics[width=0.90\linewidth]{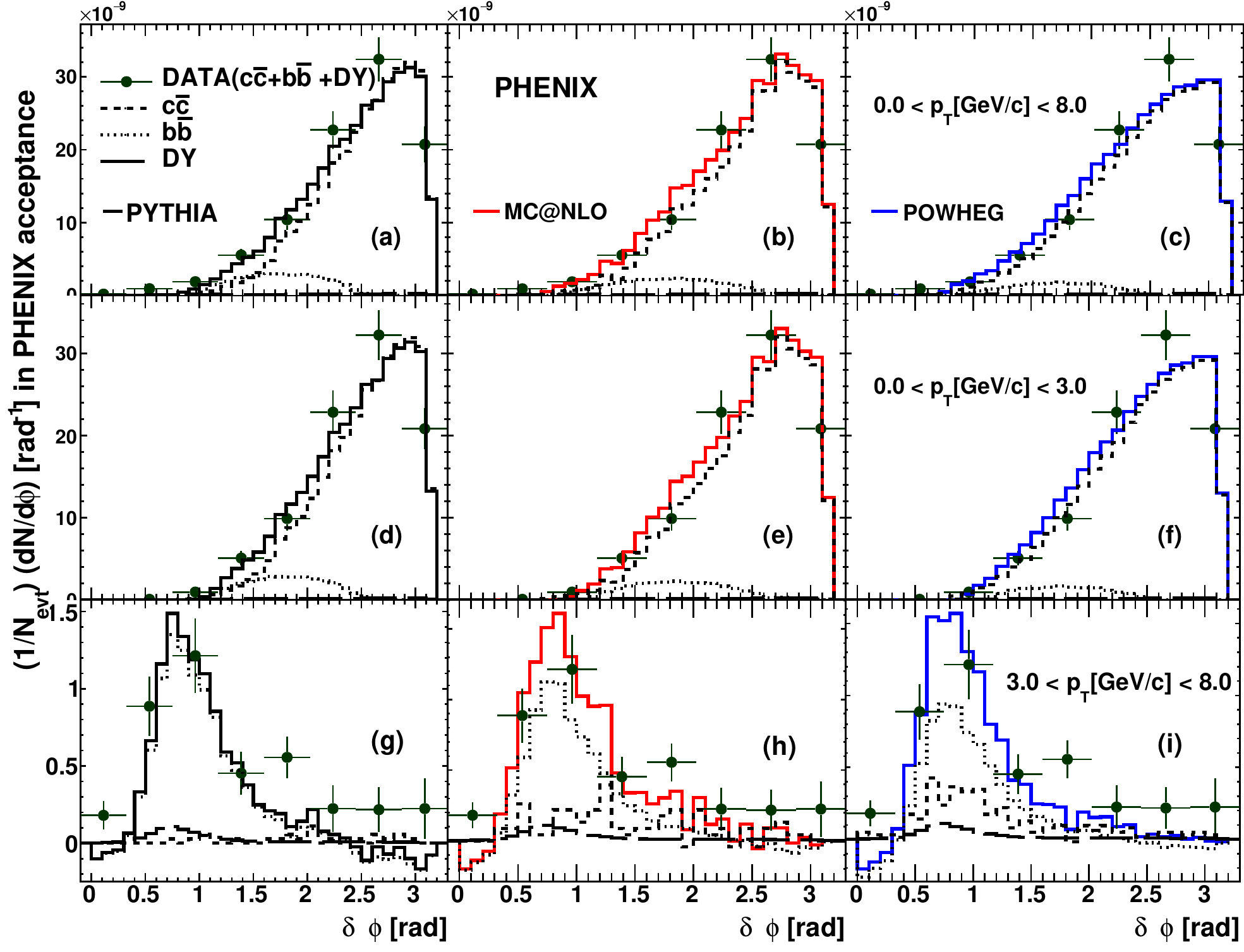}
\caption{\label{Fig:fig_deltaphi_hf}
Comparison of the $\Delta\phi$ distribution from data to the different 
MC generators. The leftmost column shows the comparison to \pythia, the 
middle column to \mcnlo, and the rightmost column shows the comparison 
to \powheg. Each row corresponds to the \pt interval indicated in the 
leftmost column. The solid line corresponds to the total HF 
contribution, dashed line represents \cc, dotted line represents \bb 
and the big dashed line shows DY contribution for a given generator. 
The normalization of different contributions is explained in the text. 
The negative yield for the simulations results from the like-sign 
subtraction performed in simulations similar to data analysis.}
\end{figure*}

\begin{figure}[!htb]
\includegraphics[width=1.0\linewidth]{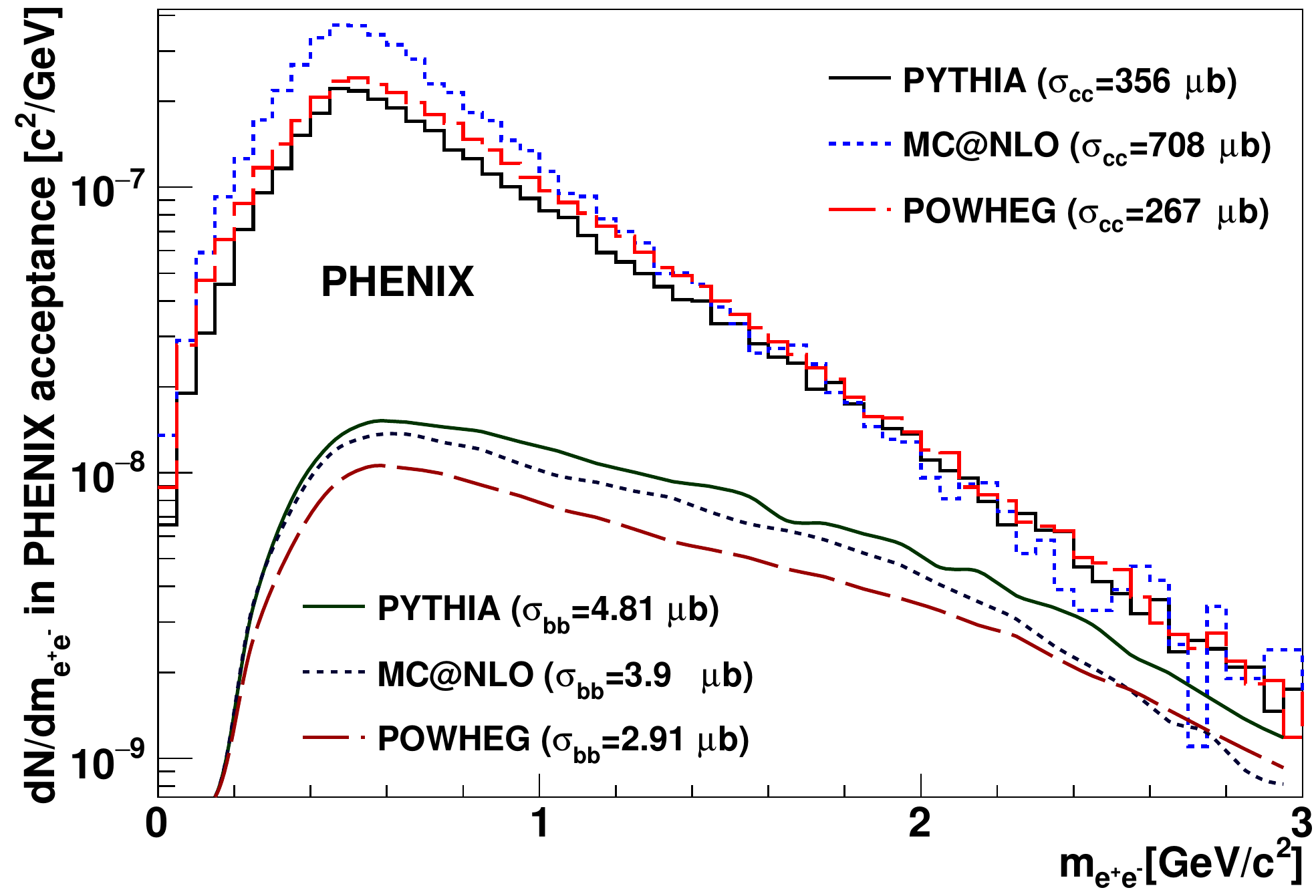}
\caption{\label{Fig:new1}
Comparison of the invariant \ee yield from \cc and \bb for \pp collisions
determined using \pythia (solid line), \mcnlo (dotted line) and \powheg
(dashed line) and normalized using the extracted cross sections.
}\end{figure}

\begin{figure}[!htb]
\includegraphics[width=0.96\linewidth]{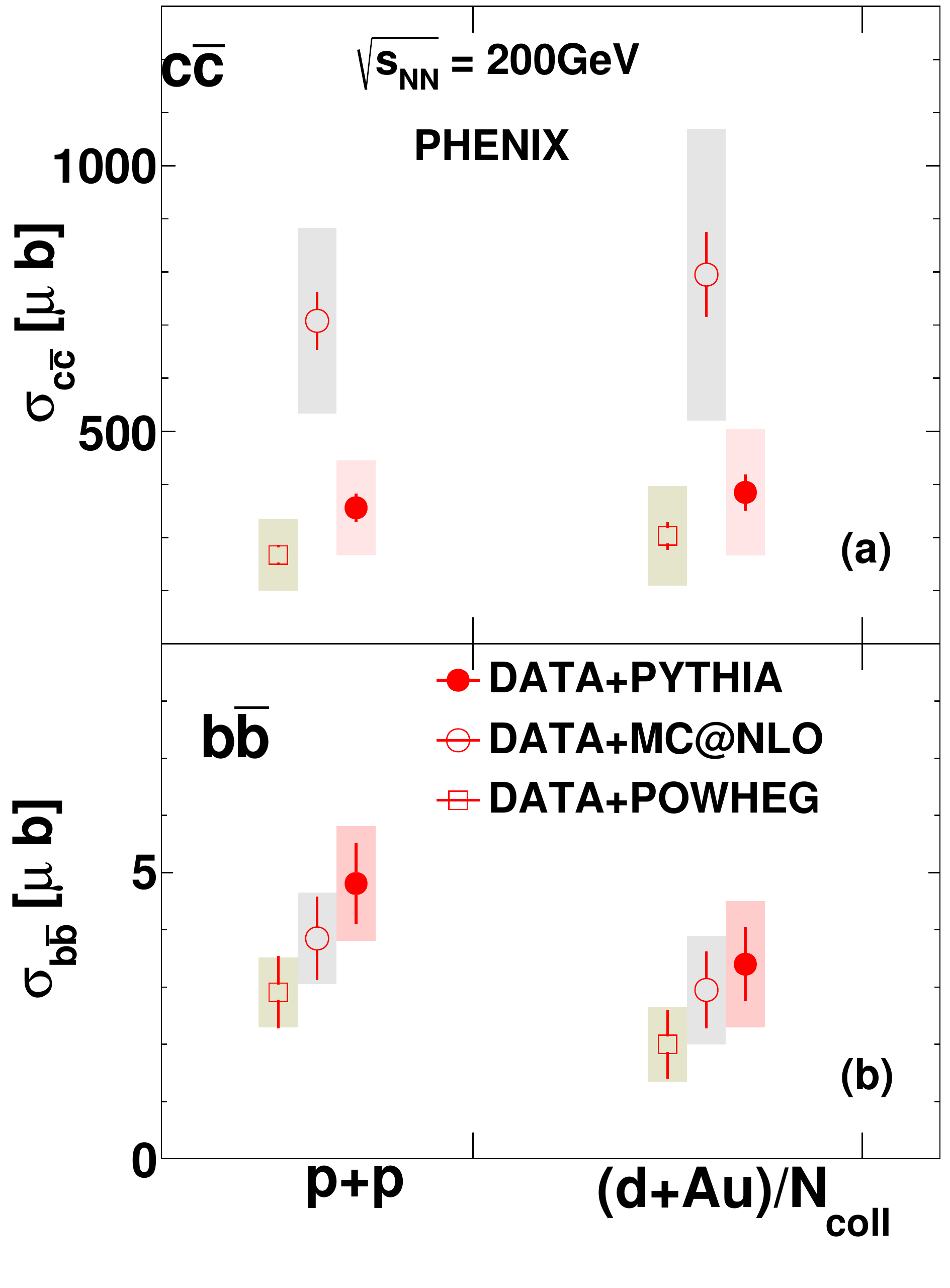}
\caption{\label{Fig:HF_xsecs_pp_dau}
The extracted 
cross sections of \cc and \bb in \pp and \dA collisions. The \dA cross-
section has been scaled down by \Ncoll to represent the equivalent nucleon-
nucleon cross section.
}
\end{figure}

\begin{table*}[hbt]
\caption{Step by step extrapolation from the number of \ee pairs for 
\mee $\ge$ 1.16 GeV/$c^2$ from \cc in the PHENIX acceptance to the 
number of \cc pairs in $4\pi$ for \pythia, \mcnlo, and \powheg. Numbers 
are in units of pairs per event using the \cc cross sections determined 
in this paper. The factors in brackets quantify the increase in number 
of pairs. We have factored out the effective branching ratio BR=0.094 
for decays of $c\rightarrow$e in the step from \ee to \cc pairs. The 
number of \cc pairs in $4\pi$ is equal to the \cc cross section in 
table~\ref{tab:xsec_pp} divided by the inelastic \pp cross section 
$\sigma_{pp}=42$mb.}
  \label{tab:ccbar}
    \begin{ruledtabular}\begin{tabular}{ccccc}
    \cc   & \pythia    & \mcnlo  & \powheg \\
    \hline
    $|y_{e^-}\&y_{e^+}|_{\rm PHENIX} ~\&\&$  & 3.20$\times$10$^{-8}$& 3.55$\times$ 10$^{-8}$ & 3.61$\times$10$^{-8}$ \\
    \mee $\ge$ 1.16 GeV/$c^2$          & & & \\         
    $|y_{e-}\&y_{e^+}|_{\rm PHENIX}$   & 1.66$\times$10$^{-7}$ (5.19) & 2.55  $\times$10$^{-7}$ (7.18) & 1.93$\times$10$^{-7}$ (5.33)  \\
    $|y_{\cc}|\le$0.5                  & 2.33$\times$10$^{-3}$ (124/BR$^2$) & 5.09 $\times$10$^{-3}$ (176.6/BR$^2$) & 1.80$\times$10$^{-3}$ (82.5/BR$^2$) \\
    $4\pi$                             & 8.48$\times$10$^{-3}$ (3.64) & 16.9 $\times$10$^{-3}$ (3.31) & 6.36$\times$10$^{-3}$ (3.53)\\
    \end{tabular}\end{ruledtabular}
\end{table*}

\begin{table*}[hbt]
\caption{ Step by step extrapolation from the number of \ee pairs for 
\mee $\ge$ 1.16 GeV/$c^2$ from \bb in the PHENIX acceptance to the 
number of \bb pairs in $4\pi$ for \pythia, \mcnlo, and \powheg. Numbers 
are in units of pairs per event using the \bb cross sections determined 
in this paper. The factors in brackets quantify the increase in number 
of pairs.  We have factored out the effective branching ratio BR=0.158
for decays of $b\rightarrow$e in the step from \ee to \bb pairs. The 
number of \bb pairs in $4\pi$ is equal to the \bb cross section in 
table~\ref{tab:xsec_pp} divided by the inelastic \pp cross section 
$\sigma_{pp}=42$mb.}
  \label{tab:bbbar}
  \centering
   \begin{ruledtabular}\begin{tabular}{ccccc}
    \bb   & \pythia    & \mcnlo  & \powheg \\
    \hline
    $|y_{e^-}\&y_{e^+}|_{\rm PHENIX}~\&\&$  & 10.3$\times$10$^{-9} $& 8.34$\times$10$^{-9}$& 6.99$\times$10$^{-9}$\\
    \mee $\ge$ 1.16 GeV/$c^2$          & & & \\         
    $|y_{e-}\&y_{e^+}|_{\rm PHENIX}$   & 2.18$\times$10$^{-8}$ (2.11) & 1.83 $\times$10$^{-8}$ (2.19)   & 1.46$\times$10$^{-8}$(2.12)   \\
    $|y_{\bb}|\le$0.5                  & 4.47$\times$10$^{-5}$ (51.1/BR$^2$) & 3.49 $\times$10$^{-5}$ (47.6/BR$^2$) & 2.61$\times$10$^{-5}$ (44.6/BR$^2$) \\
    $4\pi$                             & 11.5$\times$10$^{-5}$ (2.56) & 9.17 $\times$10$^{-5}$ (2.62) & 6.93$\times$10$^{-5}$ (2.66) \\
    \end{tabular}\end{ruledtabular}
\end{table*}

\begin{table*}
\caption{Summary of \cc and \bb cross section in \dA
collisions expressed as nucleon-nucleon equivalent cross section by 
dividing the \dA cross section by the average number of 
binary nucleon-nucleon collisions \Ncoll $=7.6\pm0.4$.} 
\label{tab:tab_xsec_dA}
\begin{ruledtabular}  \begin{tabular}{cccc}
     \dA/$N_{\rm coll}$  
& \pythia ($\mu$b)  
&  \mcnlo ($\mu$b) 
& \powheg ($\mu$b)
\\
\hline
\cc (Reanalysis) 
& 385  $\pm$ 34 (stat) $\pm$  119 (syst) 
& 795 $\pm$ 80 (stat) $\pm$ 275 (syst) 
& 303 $\pm$ 26 (stat) $\pm$ 94 (syst)
\\
\bb(Reanalysis) 
& 3.40 $\pm$ 0.65 (stat)$\pm$ 1.10 (syst) 
& 2.95 $\pm$ 0.67 (stat)$\pm$ 0.95 (syst) 
& 2.0 $\pm$ 0.6 (stat)$\pm$ 0.65 (syst)
\\
    \end{tabular}  \end{ruledtabular}
  \end{table*}

\begin{figure*}[htb]
\begin{minipage}{0.49\linewidth}
\includegraphics[width=0.99\linewidth]{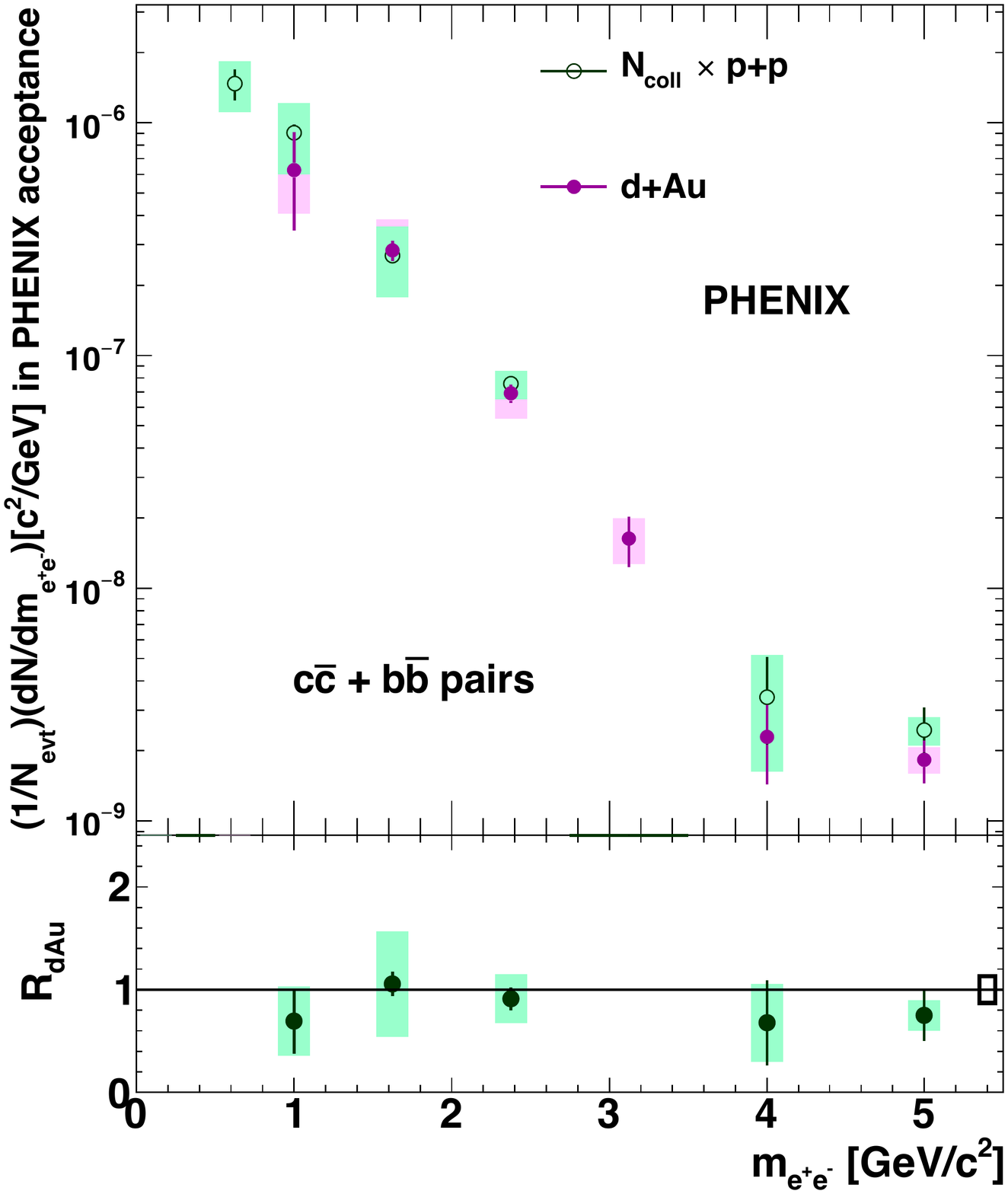}
\caption{\label{Fig:HFprojections_mass} Comparison of mass spectrum of 
\ee pairs from heavy flavor in \pp and \dA collisions. The \dA data
shown are from ~\cite{Adare:2014iwg}. The \pp yield 
has been scaled by \Ncoll $=7.6\pm0.4$ for MB \dA 
collisions.}
\end{minipage}\hfill
\begin{minipage}{0.49\linewidth}
\includegraphics[width=0.99\linewidth]{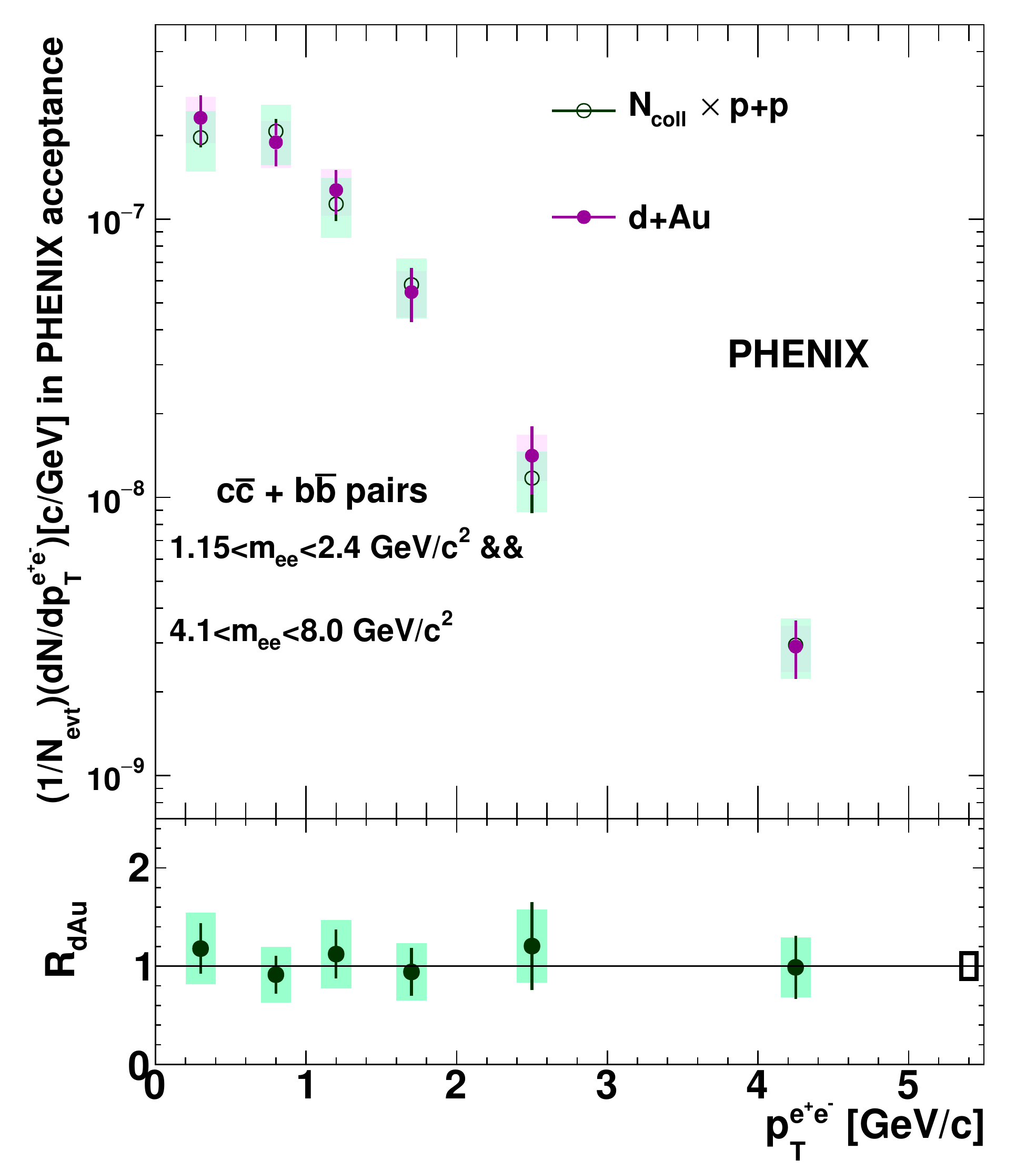}
\caption{\label{Fig:HFprojections_pt} 
Comparison of transverse momentum spectrum of \ee pairs in \pp and \dA 
collisions. The \dA data shown are from ~\cite{Adare:2014iwg},The \pp yield has been scaled by \Ncoll $=7.6\pm0.4$ for
MB \dA collisions.
}
\end{minipage}
\end{figure*}

The fitted cross sections are tabulated in Table~\ref{tab:xsec_pp}. For 
the \cc cross section we find 356, 708, and 267 $\mu$b for \pythia, 
\mcnlo and \powheg respectively. For each the statistical uncertainty 
is about 8\%, while the systematic uncertainty due to the data is 
approximately 25\%. The values cover a range of $\sim \pm 220~\mu$b 
around the average value, indicating large model dependencies that are 
further discussed in the following. The \cc cross section values are 
consistent with earlier measurements from single electron spectra that 
gave $\sigma_{c\bar{c}} = 567\pm57 (stat) \pm 244 (syst) ~\mu$b 
\cite{Adare:2006hc} and from \ee pairs that resulted in 
$\sigma_{c\bar{c}}=544{\pm}39(stat){\pm}244(syst){\pm}200^{\rm model}~\mu$b 
\cite{Adare:2008ac}. For the \bb cross section we find 
values of 4.81, 3.85, and 2.91, again for \pythia, \mcnlo, and \powheg, 
respectively. The statistical uncertainties are 15--22\% and the 
systematic uncertainties are 21\%. The observed model dependence is 
about $\sim 0.85 ~\mu$b around the average, which is significantly 
smaller than for \cc cross section.

Despite the differences between the MC generators, each one achieves an 
adequate description of the data within the uncertainties. This may be 
more easily seen in the projections of the \ee yield from heavy-flavor 
decays onto the mass and \pt axes as shown in 
Fig.~\ref{Fig:HFprojections}.

As a consistency check and to see if more discrimination power between 
the models can be achieved in terms of different projections of the 
data, we also looked at the $\Delta\phi$ distribution for \ee pairs. 
Because the analysis was done in 2 dimensions, mass and \pt, some extra 
steps were necessary. We first generated $\Delta\phi$ distributions for 
foreground and mixed unlike-sign and like-sign pairs for the mass 
region between 1.15 $<\mee<$ 2.4 GeV/$c^{2}$ and 4.1 $< \mee<$ 8.0 
GeV/$c^{2}$. The relative-acceptance corrected like-sign foreground 
$\Delta\phi$ distribution is subtracted from the unlike-sign pairs, 
which results in the $\Delta\phi$ distribution for heavy flavor pairs. 
These $\Delta\phi$ distributions were then efficiency corrected.

The data are compared to $\Delta\phi$ distributions from simulated \ee 
pairs from \cc, \bb, and Drell-Yan. For each generator, the \cc and \bb 
contributions were normalized using the cross section values from 
Table~\ref{tab:xsec_pp}.  For the \bb contribution the like-sign pairs 
were subtracted to match the procedure used in the data. The 
$\Delta\phi$ distributions from \pythia, \mcnlo, and \powheg are shown 
for different pair \pt ranges and compared to the data in 
Fig.~\ref{Fig:fig_deltaphi_hf}. Note that these distributions are for 
\ee pairs within the PHENIX acceptance. Again, all three generators 
describe the data reasonably well. The conclusions are consistent with 
those drawn from the comparison in \pt and mass. At lower \pt the yield 
is dominated by \cc production. The yield peaks at large opening angle 
$\Delta\phi$, which is characteristic for back-to-back production. At 
the same \pt, the pairs from \bb production show no pronounced 
back-to-back structure. This is consistent with the \ee pair opening 
angle being less correlated with the \bb opening angle due to the decay 
kinematics of the much heavier B mesons.  For larger \pt \bb production 
dominates, and the \ee pair opening angle $\Delta\phi$ distribution 
peaks for opening angles smaller than 90 degrees. This is due to the 
fact that these pairs result from the decay products of a single 
B-meson rather than from the \bb pair.

Only moderate differences are observed between the generators. While 
there are differences in the shape of the $\Delta\phi$ distributions 
for \cc and \bb, the main structure seen in 
Fig.~\ref{Fig:fig_deltaphi_hf} is given by the two arm detector 
acceptance. We find that the statistical significance of our data is 
insufficient to add more discriminating power between the generators by 
looking at the $\Delta\phi$ projections.

While the data are well described by all three generators within the 
PHENIX central arm acceptance and over the range they were fitted to 
the data, the obtained cross section values, tabulated in 
Table~\ref{tab:xsec_pp}, indicate that there are large systematic 
differences when extrapolated beyond the range where the models were 
fitted to the data.

The \cc cross sections found using \pythia and \powheg differ by about 
30\%, while for \mcnlo a much larger \cc cross section is determined.  
This may be due to the fact that our \powheg simulation uses the 
\pythia fragmentation scheme. Such differences can have important 
consequences if the generators are used to estimate yields from \cc 
outside the fit range, even within the PHENIX acceptance. This was 
first pointed out in \cite{Adare:2015ila} and is apparent when one 
looks at the \ee pair mass distributions below 1 GeV/$c^2$, depicted in 
Fig.~\ref{Fig:new1}. For \pythia and \powheg there is very little 
difference going from mass larger than 1 GeV/$c^2$ to zero mass, while 
for \mcnlo the \ee pair yield is much larger. This is an important 
contribution to the larger \cc cross section determined with \mcnlo.

To get a better quantitative understanding, we divided the 
extrapolation into following three steps: the first step is the 
extrapolation from the fitted \ee pairs in the PHENIX acceptance to \ee 
pairs at all masses, then to the \qq rapidity density, and finally to 
$4\pi$. These factors are tabulated in Table~\ref{tab:ccbar} and 
Table~\ref{tab:bbbar} for \cc and \bb, respectively. For \cc production 
the number of \ee pairs in the fit range is similar for \pythia, 
\mcnlo, and \powheg. This is expected, because the normalization is 
essentially fitted in the range from 1 to 2 GeV/$c^2$ where \cc 
dominates. The extrapolation to zero mass is different only for \mcnlo, 
and is responsible for about 50\% of the larger cross section for 
\mcnlo.  Going from \ee pairs in the PHENIX acceptance to the rapidity 
density $dN_{\cc}/dy$ at $y=0$ has the largest variations between 
models. The final step from \cc rapidity density to $4\pi$ has little 
model dependence indicating that the underlying rapidity distribution 
for \cc is similar in all the generators.

The situation is however different for \bb production. From 
Table~\ref{tab:bbbar} it is evident that every step of the 
extrapolation from the fit range to $4\pi$ is very similar for all 
three generators. Again this is expected because the \ee pair 
distributions from \bb production are dominated by decay kinematics 
\cite{Adare:2014iwg}. However, the number of \ee pairs in the fit range 
is different, which leads to different \bb cross section values. The 
extracted \bb cross section value using \pythia is larger as compared 
to the one derived from \mcnlo, with the latter being larger than 
\powheg.  From Figs.~\ref{Fig:doublediffsim}~and~\ref{Fig:HFprojections}, 
one can see that the shape of the \ee pair distributions from \bb 
production are very similar among the three generators. However, this 
is not the case for \ee pairs from \cc production, in particular for 
\powheg, the \ee pair momentum distribution is much harder as compared 
to other generators. Because the \cc contribution is essentially fixed in 
the mass region between 1.0 to 2.0 GeV/$c^2$ at low pair \pt, a harder 
distribution can only be accommodated in the overall fit by reducing 
\bb production, which we expect to account for all the seen variation 
between the three generators. Additional differences in the rapidity 
and momentum distribution also contribute to the very model dependent 
extrapolations of the \cc cross section in $4\pi$.

\subsection{Comparison of \pp and \dA results}

The results of the analysis of \pp data presented here can be directly 
compared to the already published \dA results \cite{Adare:2014iwg}. 
Because we are now including \powheg and are using a newer version of 
\mcnlo for the \pp analysis, we refitted the data published in 
\cite{Adare:2014iwg} with the generator versions used for \pp. We 
scaled down the \dA data by the average number of binary 
nucleon-nucleon collisions of \Ncoll($=7.6\pm0.4$). Therefore the 
resulting normalization constants represent the equivalent 
nucleon-nucleon cross section, and can be directly compared to the \pp 
results.

Table~\ref{tab:tab_xsec_dA} summarizes the \cc and \bb nucleon-nucleon 
equivalent cross sections extracted from the \dA data. We note that the 
numbers quoted here for the \mcnlo simulation are 17\% and 12\% smaller 
for \cc and \bb, respectively, compared to the numbers quoted 
in~\cite{Adare:2014iwg}. This is potentially due to using a newer \mcnlo 
version, which needed to be modified to generate charm, or a previous 
inaccuracy in how the negative weights should be used to avoid double 
counting in the \herwig fragmentation \cite{Adare:2014iwg}. In either 
case the difference is small enough to change the conclusions neither 
here nor in the original paper~\cite{Adare:2014iwg}.

\begin{figure}[hb]
  \includegraphics[width=1.0\linewidth]{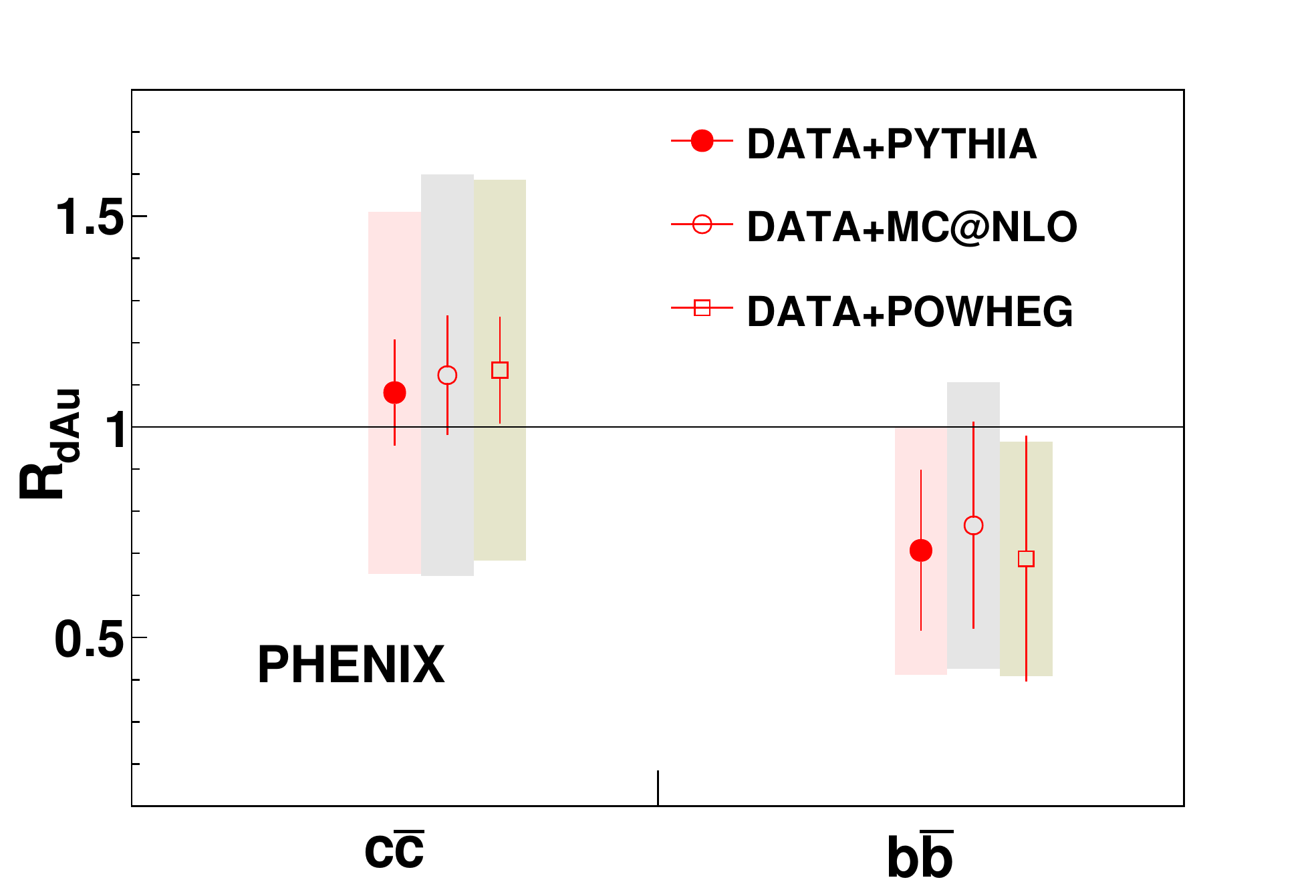}
\caption{The nuclear modification factor \Rda of \cc and \bb pairs 
constructed using the cross sections. The \dA cross sections are scaled 
down by \Ncoll $=7.6\pm0.4$.
}
\label{Fig:fig_rda_bb_cc}
\end{figure}

The comparison of the numbers in Table~\ref{tab:xsec_pp} and 
Table~\ref{tab:tab_xsec_dA} is shown graphically in 
Fig.~\ref{Fig:HF_xsecs_pp_dau}.  We see the same model dependence for 
\dA as was seen for \pp. For a given model, the obtained \cc cross 
sections are consistent within the given uncertainties in \pp and \dA. 
We also looked at the ratio (or nuclear modification) of cross sections 
of \cc and \bb in \dA and \pp and this is plotted in 
Fig.~\ref{Fig:fig_rda_bb_cc}. This ratio is similar for all the event 
generators and no deviation from unity is observed.

Fig.~\ref{Fig:HFprojections_mass} and Fig.~\ref{Fig:HFprojections_pt} 
show a direct comparison of the measured mass and \pt spectra of \ee 
pairs from heavy flavor decays between \pp and \dA systems. The top 
panels show an overlay of mass and \pt spectra in \pp and MB 
\dA collisions, where we scaled the \pp yield by \Ncoll ($=7.6\pm0.4$), 
corresponding to MB \dA collisions. Within the statistical 
precision of the data, the mass and \pt spectra in \pp and \dA agree 
with each other. The bottom panel in these figures show the ratio of 
\dA to \pp data. Given the uncertainties, the ratios are consistent 
with 1.  While the \ee pair data shows no evidence for any nuclear 
modification to the \cc and \bb production, due to the large 
statistical and systematic uncertainty, they would not be sensitive to 
effects smaller than 30\%. For example, the observed modification of 
single electron spectra seen in \dA collisions \cite{Adare:2012yxa} 
could result in a change of 30\% in the \ee pair mass and \pt 
distributions, but that might not be seen here due to the large 
uncertainties.

\section{Summary and conclusions}

We present \ee pair measurements from heavy flavor decays in \pp 
collisions at $\sqrt{s}=200$ GeV. The data are shown multi-differential 
as a function of pair mass, ~\pt, and $\Delta\phi$. By comparing the 
\ee pair data to pQCD calculations, the \cc and \bb production cross 
sections can be constrained. Three different pQCD based Monte-Carlo 
models are used: \pythia, \mcnlo, and \powheg. We find that the \cc 
production cross section ranges from 267 to 708 $\mu$b with a 
statistical (systematic) uncertainty of about 8\% (25\%). The \bb 
production cross section ranges from 2.9 to 4.8 $\mu$b with a 
statistical (systematic) uncertainty of 15--22\% (21\%).

The \ee pair distributions obtained from \pythia, \mcnlo, and \powheg 
within the PHENIX acceptance, once normalized to data, were found to be 
consistent in mass, \pt and $\Delta\phi$. In case of \cc, the 
extrapolation beyond the measured range shows substantial model 
dependence. This is evident by more than 400 $\mu$b difference between 
the obtained \cc cross sections, which is more than 100\% compared to 
the average value.

We find a smaller variation for \bb, which is less than 50\% of the 
average \bb cross section value. This variation is entirely due to the 
model dependence of \cc production. The extrapolation of \bb from the 
measured range shows little model dependence, because in our acceptance 
the decay kinematics dominate the \ee pair distributions from \bb.

We compare our \pp results directly to \ee pair measurements from 
MB \dA collisions. The \cc and \bb cross sections are 
determined in the same way for both the systems. Although there is 
significant model dependence in extracting the cross sections, within a 
given model, there is no difference between the cross sections 
determined from \pp and the equivalent nucleon-nucleon cross section 
obtained from \dA . Furthermore, we compare directly the measured \ee 
pair mass and \pt distributions from \pp and \dA. After scaling with 
the number of binary collisions, we observe no evidence for nuclear 
modifications of heavy flavor production in the \dA system within our 
experimental uncertainties.

\section*{ACKNOWLEDGMENTS}   

We thank the staff of the Collider-Accelerator and Physics
Departments at Brookhaven National Laboratory and the staff of
the other PHENIX participating institutions for their vital
contributions.  We acknowledge support from the 
Office of Nuclear Physics in the
Office of Science of the Department of Energy,
the National Science Foundation, 
a sponsored research grant from Renaissance Technologies LLC,
Abilene Christian University Research Council, 
Research Foundation of SUNY, and
Dean of the College of Arts and Sciences, Vanderbilt University 
(U.S.A),
Ministry of Education, Culture, Sports, Science, and Technology
and the Japan Society for the Promotion of Science (Japan),
Conselho Nacional de Desenvolvimento Cient\'{\i}fico e
Tecnol{\'o}gico and Funda\c c{\~a}o de Amparo {\`a} Pesquisa do
Estado de S{\~a}o Paulo (Brazil),
Natural Science Foundation of China (People's Republic of China),
Croatian Science Foundation and
Ministry of Science and Education (Croatia),
Ministry of Education, Youth and Sports (Czech Republic),
Centre National de la Recherche Scientifique, Commissariat
{\`a} l'{\'E}nergie Atomique, and Institut National de Physique
Nucl{\'e}aire et de Physique des Particules (France),
Bundesministerium f\"ur Bildung und Forschung, Deutscher
Akademischer Austausch Dienst, and Alexander von Humboldt Stiftung (Germany),
National Science Fund, OTKA, EFOP, and the Ch. Simonyi Fund (Hungary),
Department of Atomic Energy and Department of Science and Technology (India), 
Israel Science Foundation (Israel), 
Basic Science Research Program through NRF of the Ministry of Education (Korea),
Physics Department, Lahore University of Management Sciences (Pakistan),
Ministry of Education and Science, Russian Academy of Sciences,
Federal Agency of Atomic Energy (Russia),
VR and Wallenberg Foundation (Sweden), 
the U.S. Civilian Research and Development Foundation for the
Independent States of the Former Soviet Union, 
the Hungarian American Enterprise Scholarship Fund,
and the US-Israel Binational Science Foundation.


%
 
\end{document}